\documentclass[twocolumn,amsmath,amssymb,aps,prb,longbibliography]{revtex4-1}
\usepackage[latin9]{inputenc}
\usepackage{amsmath}
\usepackage{amssymb}
\usepackage{graphicx}
\usepackage{nicefrac}
\usepackage{esint}

\makeatletter
\@ifundefined{textcolor}{}
{%
 \definecolor{BLACK}{gray}{0}
 \definecolor{WHITE}{gray}{1}
 \definecolor{RED}{rgb}{1,0,0}
 \definecolor{GREEN}{rgb}{0,1,0}
 \definecolor{BLUE}{rgb}{0,0,1}
 \definecolor{CYAN}{cmyk}{1,0,0,0}
 \definecolor{MAGENTA}{cmyk}{0,1,0,0}
 \definecolor{YELLOW}{cmyk}{0,0,1,0}
}

\usepackage{float}

\makeatother

\begin{document}

\title{Decoherence of charge density waves in beam splitters for interacting quantum wires}

\author{Andreas Schulz, Imke Schneider, James Anglin}

\affiliation{Department of Physics and Research Center OPTIMAS, University
of Kaiserslautern, 67663 Kaiserslautern, Germany}
\begin{abstract}
Simple intersections between one-dimensional
channels can act as coherent beam splitters for non-interacting electrons.
Here we examine how coherent splitting at such intersections
is affected by inter-particle interactions, in the special case of an intersection of topological edge states. We  derive an effective impurity model which represents the edge-state intersection
within Luttinger liquid theory at low energy. For Luttinger $K=\nicefrac{1}{2}$, we compute the exact time-dependent
expectation values of the charge density as well as the density-density correlation
functions. In general a single incoming charge density wave packet will split into four outgoing wave packets with transmission and reflection coefficients depending on the strengths of the tunnelling processes between the wires at the junction. We find that when multiple charge density wave packets from different directions
pass through the intersection at the same time, reflection and splitting
of the packets depend on the relative phases of the waves. Active use of this phase-dependent splitting of wave packets may make Luttinger interferometry possible. We also find that coherent incident packets generally suffer
partial decoherence from the intersection, with some of their initially coherent
signal being transferred into correlated quantum noise. In an extreme case four incident coherent wave packets can be transformed entirely into density-density correlations, with the charge density itself having zero expectation value everywhere in the final state. 
\end{abstract}
\maketitle

\section{Introduction}

Interacting electrons within nanoscopic one-dimensional metals are
well described by Haldane's Luttinger liquid theory in terms of bosonic
excitations of charge density waves \cite{Tomonaga1950,Luttinger1963, Haldane1981a, Giamarchi2004}. The bosonic collective modes---Luttinger plasmons---are non-interacting at low energies and propagate linearly without dispersion. In this sense they behave much like photons.   
Just as photons may be generated by time-dependent
charge distributions and detected by the charge movements that they
induce when they are absorbed, Luttinger plasmons may be generated
and detected via external electromagnetic fields. Quasi-monochromatic coherent
wave packets of plasmons can be transmitted through quantum wires in much
the same way that coherent laser pulses of photons can be through optical fibers \cite{Hashisaka2017, Kamata2013}.

Among the most powerful technological applications of lasers is interferometry,
which is enabled by the coherent splitting and recombining of light
beams in linear beam splitters. Single-electron wave packets can similarly
split and combine coherently if they propagate in one-dimensional
channels which intersect. While in general electrons encountering
a junction of two quantum wires may be transmitted in all three outgoing
directions and possibly also reflected back along their incident wire,
it has recently been shown that intersections of topologically protected
edge state channels can act just like optical beam splitters 
for non-interacting
electrons, splitting incoming packets into exactly two outgoing packets with zero reflection \cite{Qiao2014,Anglin2017}. Whether for
this conveniently light-like splitting or for more general multi-way
splitting, however, the question of signal coherence through quantum
wire junctions depends for real electrons on the effects of Coulomb
interactions.

Research on transport in inhomogeneous interacting  quantum wires  has a long history. It was established early on that even the smallest impurity potential in an effectively one-dimensional wire with repulsive interactions can introduce dramatic effects due to backward scattering which can block electric \cite{Kane1992a,Kane1992,Furusaki1993,Furusaki1996,Yue1994} or magnetic \cite{Eggert1992,Pereira2004} conductance at low temperatures. More general inhomogeneities have then been considered in order to describe the coupling of a wire to leads \cite{Safi1995, Safi1999,Enss2005, Janzen2006}.  Remarkably, for  sharp junctions of two wire regions with different effective band widths, chemical potentials and interaction strengths, a line of perfectly conducting fixed points exists where the backscattering at the junction vanishes despite the inhomogeneity of the system \cite{Sedlmayr2012,Sedlmayr2014,Morath2016}.  Furthermore, correlation functions and the local  density of states near inhomogeneities and boundaries are now understood well enough to establish that inhomogeneities induce wave-like modulations in the local density of states, and that there is a sharp reduction of the density of states near boundaries \cite{Mattsson1997,Eggert2000, Schneider2008, Schneider2010, Schuricht2012, Soeffing2013}.

Investigations of the real-time dynamics of charge transport in one-dimensional wires and the partitioning and recombining of  coherently propagating wave-packets have opened the field of electron quantum optics.  Of particular interest currently are on-demand single electron sources  \cite{Ivanov1997, Levitov1996, Keeling2006,Bauerle2018}, which have been experimentally realized using periodically driven mesoscopic capacitors \cite{Feve2007,Mahe2010} or  time-dependent voltages \cite{Dubois2013}.  
The interference of two such single charges has been probed when the two excitations collide at an intersection in a Hong-Ou-Mandel setup \cite{Hong1987, Beugnon2006,Wahl2014}. 

In all such problems of excitations passing through intersections, the interaction-induced relaxation
and decoherence mechanisms after the initial creation of the excitation are  important to understand. 
Generally speaking, a sufficiently narrow intersection of quantum wires may behave as
a linear element for electrons, but the evolution of interacting electrons
along the length of a quantum wire is extremely nonlinear. For the
quasi-particles whose propagation in quantum wires is conveniently
linear, the Luttinger plasmons, an intersection of two wires is on the other
hand a highly nonlinear impurity. When interacting electrons pass through an intersection impurity, therefore, nonlinear effects are generic. Their influence on the propagation and decoherence of charge density wave packets will be the subject of this paper.

We will find that the intersection can mix and redirect multiple incident charge
density wave packets of matching frequency in a manner that depends
on the relative phases among the incident packets. The effect of the
relative incident phases on the outgoing signals is not simply to redistribute
their intensities in different emission directions, however, as in an
optical beam splitter, but also to make the transmitted signals
more or less coherent: the coherent incident charge density waves
are partially converted into correlated quantum noise. This effect may be a limitation on Luttinger
interferometers that would operate exactly like optical ones, but
it may also offer a new way to measure relative phases, with a new kind of interferometer based on Luttinger
liquids.

Our paper is organized as follows. We begin in Section II with our
basic model of  two one-dimensional fermionic channels that effectively intersect at a point. Motivated by a recent analytical result for non-interacting fermions in intersecting topological edge states \cite{Anglin2017}, we will adopt a simple but realizable model in which the intersection lets fermions tunnel between channels but no in-wire backscattering is induced.

In Section III we then introduce short-ranged screened Coulomb interactions
among our fermions and use the one-loop renormalization group to derive an effective low-energy theory for the
interacting system. Because of the particular kind of single-particle tunnelling which our intersection allows, we will find that in the low-energy limit of the interacting system the intersection is described by just one particular two-body term. 

In Section IV we will then focus on one informative case of our low-energy theory, namely the special symmetry point of Luttinger parameter $K=1/2$, which can be solved exactly by refermionization. We will compute the time- and space-dependent expectation values of the charge density, as well as
its two-point correlation functions, for initial quantum states with charge density wave packets
incident on the intersection. These will be the main results of our paper, showing how the intersection's nonlinear action on Luttinger plasmons produces both phase-sensitive transmission and decoherence. 

In Section V we will prove that passage of excitations through the intersection can create long-range quantum entanglement between fermionic degrees of freedom, by applying the Peres-Horodecki non-separability criterion in a two-qubit subspace of the many-body Hilbert space. In Section VI we will conclude with a brief discussion of how our intersection may be considered as a nonlinear beam splitter. Two appendices will then review technical details that may be unfamiliar to some readers.

\section{Intersection of Two Quantum Wires}

\begin{figure}
\center\includegraphics[width=0.9\columnwidth]{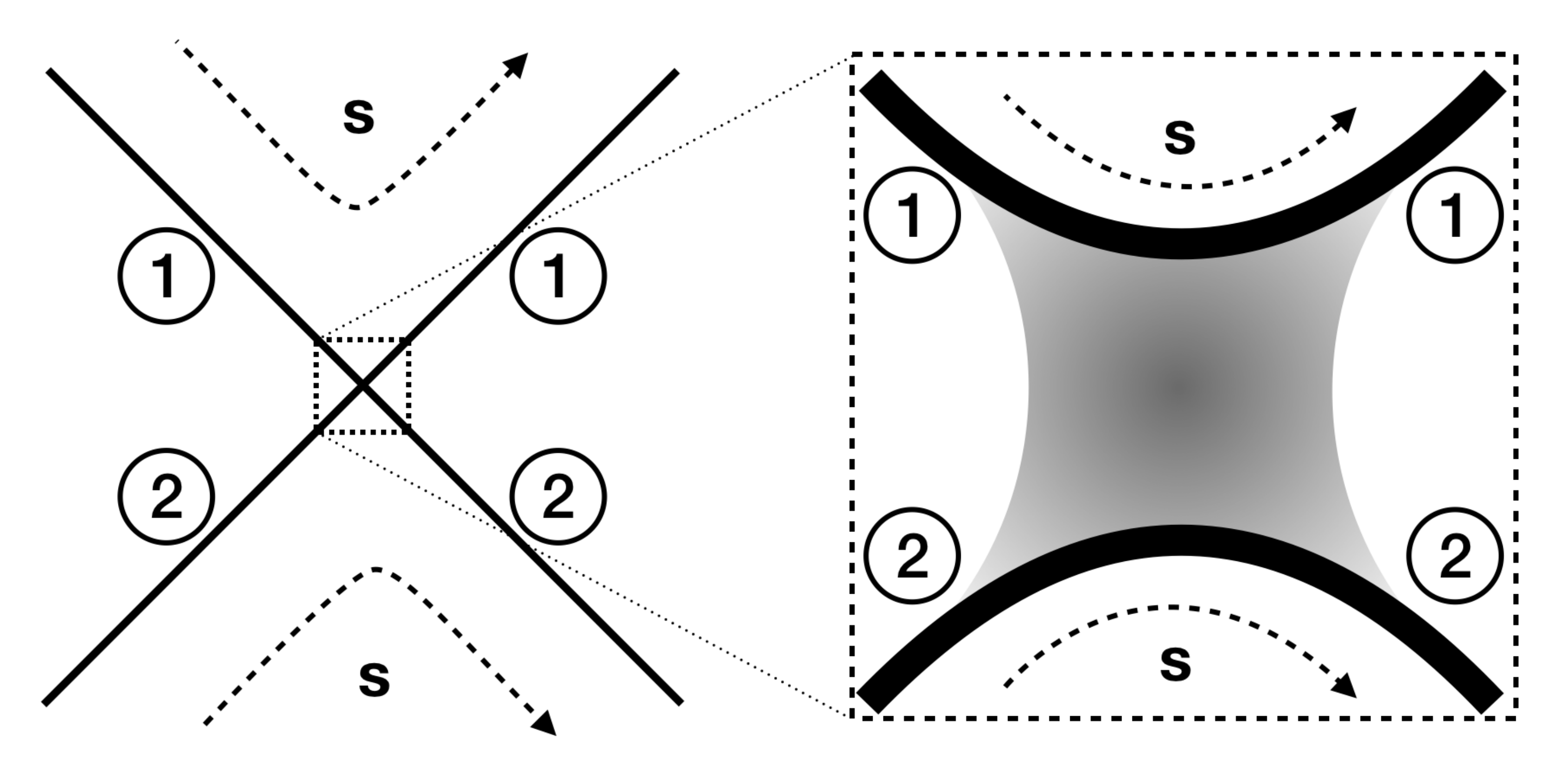}\caption{\label{fig:Intersection}Schematic representation of the intersection.
Two one-dimensional quantum wires effectively intersect at a common
point (left panel), although microscopically (right panel) the `intersection'
may really only be a close approach which allows particles to tunnel
between the wires within a small region. The continuous wires labelled
1 and 2 are therefore the `northwest-to-northeast' and `southwest-to-southeast' angled
lines, as indicated. The coordinate $s$ within each wire is chosen
to run in the directions indicated by the bent dashed lines with arrowheads.
Fermions within each wire can move in both positive and negative $s$
directions. Those moving in the positive $s$ direction in each wire will be denoted as right-movers with
`R' subscripts, while left-movers going in the negative $s$ directions will have `L' subscripts.}
\end{figure}

We consider two quantum wires ($j=1,2$) occupied by spinless one-dimensional
fermions. We assume that in each wire the fermions' dispersion relation can be taken, within
the entire range of frequencies that is relevant to our discussion,
to consist of two mirror-symmetrical linear regions around the Fermi level $k=\pm k_{F}$,
so that with our two wires we effectively have two species of right-moving fermions, of which the associated operators will be labelled with $R$ indices, and two species of left-moving fermions, whose operators are distinguished with $L$ indices.

As is typical in one-dimensional many-body theory, all of our analysis will depend crucially on the linearity of our fermionic dispersion relation. 
Implicitly, therefore, whenever we write any quantum field operator $\hat{\psi}(s)$ with spatial position argument $s$, we really mean that $s$ is sufficiently smeared to project this field operator into the space of many-body quantum states whose excited particles are all of sufficiently long wavelength to have linear dispersion. This is a basic issue in one-dimensional many-body physics, discussed in standard works \cite{Gogolin1998,Giamarchi2004}. Its important implications for our representation of charge density waves are explained in Appendix \ref{AppendixB}. 

We consider our two wires to be far
apart from each other everywhere except within a small region in which
they approach each other closely enough for particles to tunnel from
one wire to the other; see Fig.~1. The region of close proximity
is to be small enough compared to all excitation wavelengths that
it can be regarded as effectively point-like; we assign this effective
point the same co-ordinate $s=0$ along both wires, and refer to the
point $s=0$ as `the intersection', even though our wires may
not literally cross. Our convention within each wire is that the right-moving
particles are those that move in the direction of increasing $s$. 

To avoid writing too many separate equations for $L$ and $R$ fields, we define $\mp_{\alpha}$ to be $-1$ for $\alpha=R$ and $1$ for $\alpha=L$. The Hamiltonian representing the single-particle dynamics of the system
(\textit{i.e.} without interactions among the fermions) can then be written as 
\begin{eqnarray}\label{eq:H_single_particle_fermion}
\hat{H}_{1P} & = & \sum_{\alpha=L,R}\sum_{j=1,2}\int ds\,\Big\{\hat{\psi}_{\alpha j}^{\dagger}\left(\mp_{\alpha}i v_{F}\partial_{s}\right)\hat{\psi}_{\alpha j}\Big\}\nonumber\\
 &&\qquad + g_{F}\hat{t}_{F} + g_{B}\hat{t}_{B}\nonumber\\[6pt]
 \hat{t}_{F}&=& \hat{\psi}_{R1}^{\dagger}(0)\hat{\psi}_{R2}(0) +\hat{\psi}_{L1}^{\dagger}(0)\hat{\psi}_{L2}(0) + \mathrm{H.c.}\nonumber\\
 \hat{t}_{B}&=& \hat{\psi}_{R1}^{\dagger}(0)\hat{\psi}_{L2}(0) +\hat{\psi}_{L1}^{\dagger}(0)\hat{\psi}_{R2}(0) + \mathrm{H.c.}\;,\end{eqnarray}
where $g_{F,B}$ are tunnelling strength coefficients. There are two distinct coefficients $g_{F,B}$ because if either a
right- or left-moving fermion should tunnel from one wire into the
other wire, then it may thereafter move in either direction
along the second wire---and the rates of tunnelling may not be the same
for these two directional cases. We designate tunnelling which keeps
the particle moving in the same $s$-direction in the new wire as
`forward' (F), while the process is `backward' (B) if the direction
of motion in $s$ is reversed after tunnelling. Note that we assume that the wires have sufficient microscopic left-right symmetry in the intersection region that we may consider $g_{F,B}$ to be the same for right-movers and left-movers.

In this paper we will initially allow both $g_{F}$ and $g_{B}$ to be non-zero, but
we will have in mind $g_{B}\gg g_{F}$, and our explicit results will effectively be for $g_{F}=0$.
This simplification is motivated by the concrete example of one-dimensional channels that are a particular kind of topological edge state, namely the quantum valley
Hall (QVH) zero-line modes (ZLM) in a right-angled topological intersection
in graphene, for which the two tunnelling coefficients (as here defined)
turn out for topological reasons to be $g_{B}=v_{F}^{-1}\pi/4$ and
$g_{F}=0$ exactly \cite{Anglin2017}. This is indeed somewhat counter-intuitive: fermions that encounter the four-way ZLM intersection can turn to both left and right, but cannot proceed directly ahead. With our convention for the direction of $s$ in the two wires, it is thus the `backward' processes which are favored while the `forward' processes are suppressed. The reasons for this are topological.

A more important simplification that occurs in the QVH ZLM case \cite{Anglin2017}, also for topological reasons, is that the intersection produces no reflection within the same wire. Both tunnelling terms $\hat{t}_{F,B}$ that are included in $\hat{H}_{1P}$ move fermions from one wire to the other. Generalizations to include in-wire reflection
from the intersection, or to allow $g_{F}\sim g_{B}$, are considered in Appendix B. There we show that although in-wire reflection will effectively `cut' the two wires for sufficiently long wavelength excitations, for small enough microscopic in-wire reflection there will be a wide range of long wavelengths within which the effects of in-wire reflection remain negligible.

We also consider two-particle interactions: short-ranged screened
Coulomb repulsion between fermions along the length of each wire,
as well as localized two-particle tunnelling processes at the
intersection. We therefore take the total Hamiltonian to be  
$\hat{H}=\hat{H}_{1P}+\hat{H}_\mathrm{int}$,
for (following the standard $g$-ology notation \cite{Giamarchi2004})
\begin{eqnarray}\label{eq:H_Interaction_fermionic}
\hat{H}_\mathrm{int} & = & \sum_{j}\,\int ds\left\{\frac{g_4}{2}\left(\hat{\rho}_{R j}\hat{\rho}_{R j}+\hat{\rho}_{L j}\hat{\rho}_{L j} \right) + g_2 \hat{\rho}_{R j}\hat{\rho}_{L j}  \right\}\nonumber \\
&&   + 
 G_{F}\hat{T}_{F} + G_{B}\hat{T}_{B}  \nonumber \\[6pt]
\hat{T}_{F}&=& \hat{\psi}^{\dagger}_{R1}(0)\hat{\psi}^{\dagger}_{L2}(0)\hat{\psi}_{R2}(0)\hat{\psi}_{L1}(0)+\mathrm{H.c.}\nonumber\\
\hat{T}_{B}&=& \hat{\psi}^{\dagger}_{L1}(0)\hat{\psi}^{\dagger}_{L2}(0)\hat{\psi}_{R1}(0)\hat{\psi}_{R2}(0)+\mathrm{H.c.}\;,\end{eqnarray}
where $\hat{\rho}_{R j}=\hat{\psi}^{\dagger}_{R j}\hat{\psi}_{R j}$ and  $\hat{\rho}_{L j}=\hat{\psi}^{\dagger}_{L j}\hat{\psi}_{L j}$ denote the densities of right and left movers, respectively.  All operators are normal ordered with respect to
the non-interacting Dirac sea.
The two new
$\hat{T}_{F,B}$ terms at the intersection provide simultaneous
tunnelling between the wires by two particles at once. Several other
two-body impurity terms at the intersection can exist besides these
particular two, but we will see below that these are the only ones
which are \emph{relevant} in the renormalization group sense. As the
sketch Fig.~2 indicates, both these relevant terms effectively involve
two particles passing through each other as they tunnel between wires
in opposite directions. The tunnelling strengths $G_{B}$ and $G_{F}$
apply respectively to processes where the particles pass each other
while `turning', as in Figs.~2a) and 2b), or instead pass each other
going straight through the intersection, as in Figs.~2c) and 2d). Our renormalization group analysis in the next Section will show that when the microscopic single-fermion tunnelling is sufficiently suppressed in the forward direction, then at long wavelengths the forward-type two-fermion tunnelling will have $G_{F}$ negligible as well, allowing us to consider simple cases with only $\hat{T}_{B}$ terms in the effective long-wavelength Hamiltonian.

\begin{figure}
\center\includegraphics[width=0.9\columnwidth]{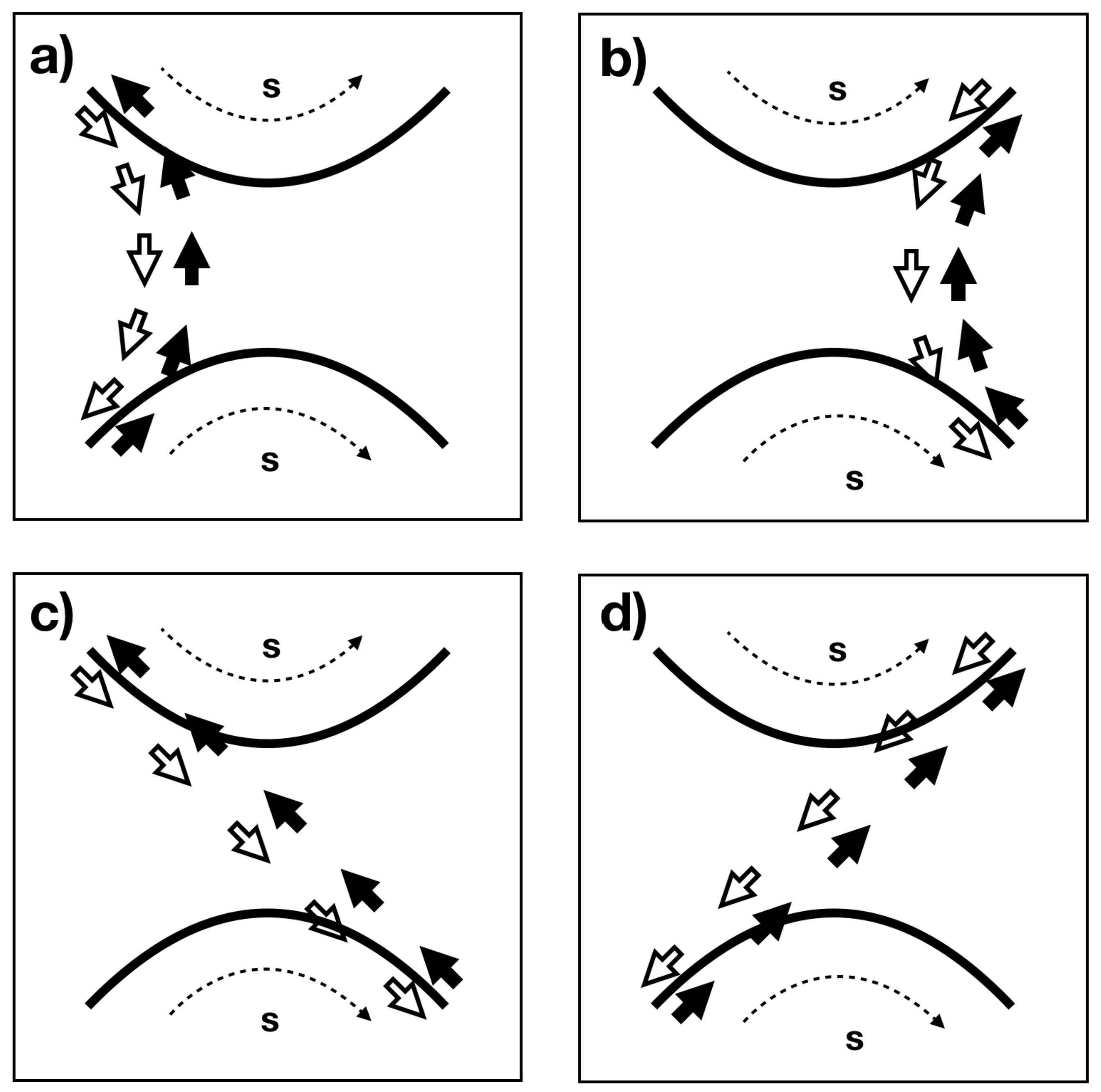}
\caption{\label{fig:Processes}Relevant two-particle tunnelling processes at
the intersection. Thick black curves denote the two wires; wide arrows
indicate simultaneous tunnelling of two particles from one wire to
the other. \\
\textbf{Upper
panels:} the two conjugate processes in $\hat{T}_{B}$.\\
\textbf{Lower panels:} the two conjugate processes in $\hat{T}_{F}$.}
\end{figure}

For a sufficiently small intersection region, both $G_{F}$ and $G_{B}$
can be small. This does not mean that the effect of interparticle
interactions on propagation through the intersection is small, however,
because the bulk interactions are always important for long-wavelength
excitations. For bulk two-body interaction strength $V_{0}>0$, the excitations which propagate freely
down the wires are not individual fermions, but collective Luttinger
plasmons. Even if the microscopic two-body tunnelling strengths at the intersection $G_{F}$ and $G_{B}$ are small,
interactions will still affect how
excitations pass through the intersection, because interactions will determine what these excitations actually are.

To determine the effect of the intersection, therefore, we will
translate our system of interacting fermions into the equivalent bosonized
model of Luttinger plasmons, which are non-interacting and dispersionless
along the length of the wires. In terms of plasmons, the simple one- and two-body fermion operators $\hat{t}_{F,B}$ and $\hat{T}_{F,B}$ will be extremely nonlinear. To then uncover
the effect of these nonlinear terms, we will use the renormalization
group (RG) to derive a simpler approximation to them which will be
valid for low-energy excitations.

\section{Effective Low-Energy Theory}

Our work in this Section is inspired by similar studies on crossed nanotubes by Komnik and Egger
\cite{Komnik1998,Komnik2001}, and refines recent results in the context
of a  bilayer graphene systems \cite{Wieder2015,Teo2009}.

\subsection{Bosonization}

Within an Abelian bosonization framework \cite{Giamarchi2004, Gogolin1998}
we can map our fermion operators onto bosonic
fields by defining
\begin{eqnarray}\label{eq:Bosonization_identity_rescaled}
\hat{\psi}_{\alpha j}(s) & = & \frac{\eta_{\alpha j}}{\sqrt{2\pi\,\lambda}}e^{i\left(\pm_{\alpha}\sqrt{\pi K}\,\hat{\Phi}_{j}(s)-\sqrt{\pi/K}\,\hat{\Theta}_{j}(s)\right)}\;,
\end{eqnarray}
where $\pm_{\alpha}$ is $+$ for $\alpha=R$, $-$ for $\alpha=L$. Here $\lambda$ is the ultraviolet cut-off length scale representing the extent of the $k$-space window around $k_{F}$ in which the fermion dispersion relation is linear, allowing bosonization to work. The Klein factors $\eta_{\alpha j}^{\dagger}=\eta_{\alpha j}$ ensure anti-commutation relations among different branches and
obey a Clifford algebra $\left\{ \eta_{\alpha j},\eta_{\alpha' j'}\right\} =2\delta_{\alpha \alpha'}\delta_{jj'}$ for $\alpha,\alpha' = R,L$ and $j,j'=1,2$.
 In particular we follow Komnik and Egger \cite{Komnik1998} in identifying the products
of Klein factors with Pauli matrices:  
\begin{align}
\eta_{\alpha1}\eta_{\alpha2}&=i\sigma_{x},\quad &\eta_{\alpha1}\eta_{-\alpha2}&=\mp_\alpha i\sigma_{y},\nonumber \\
\eta_{\alpha1}\eta_{-\alpha 1}&=-\mp_\alpha i\sigma_{z},\quad &\eta_{\alpha2}\eta_{-\alpha2}&=\mp_\alpha i\sigma_{z}\enspace. \nonumber
\end{align}
The bosonic fields and their duals  satisfy $\left[\hat{\Phi}_{j}(s),\hat{\Theta}_{j'}(s')\right]=-\frac{i}{2}\delta_{jj'}\textrm{sgn}(s-s')$. They directly measure the collective low-energy density modes of the quantum fluid. For later convenience we additionally introduce  left- and right-moving fields such that $\hat{\Phi}_{j}=\hat{\Phi}_{Rj}+\hat{\Phi}_{Lj}$ and  $\hat{\Theta}_{j}=\hat{\Phi}_{Lj}-\hat{\Phi}_{Rj}$.

The bulk interaction parameter $K$ encodes the two-particle Coulomb interaction
in the first line of Eqn.~(\ref{eq:H_Interaction_fermionic}). 
We assume the two wires are sufficiently similar within the region of interest to have identical Luttinger parameters
$K=\sqrt{\frac{1+\frac{g_4}{2 \pi v_{F}}-\frac{g_2}{2 \pi v_{F}}}{1+\frac{g_4}{2\pi v_{F}}+\frac{g_2}{2\pi v_{F}}}}$ for both wires, as
well as identical bare plasmon velocities $u_{1}=u_{2}=u\equiv v_{F} \sqrt{\left(1+\frac{g_4}{2 \pi v_{F}}\right)^2-\left(\frac{g_2}{2 \pi v_{F}}\right)^2}$. We then choose units in which $u=v_{F}/K=1$ assuming $g_2=g_4$. As usual, products of $\hat{\psi}_{\alpha j}^\dagger$ and  $\hat{\psi}_{\alpha j}$ are to be understood in terms of a
standard point splitting and normal ordering prescription which yields the following bosonized expressions for the densities in each wire $j=1,2$:  
\begin{eqnarray}
\hat{\rho}_{R j}+\hat{\rho}_{L j}&=&\sqrt{\frac{K}{\pi}}\partial_s \hat{\Phi}_j\\
\hat{\rho}_{L j}-\hat{\rho}_{R j}&=&\sqrt{\frac{1}{\pi K}}\partial_s \hat{\Theta}_j\enspace.
\end{eqnarray} 
 Integrating over the densities gives the total charge operator
\begin{eqnarray}\label{chargeoperator}
\hat{q}_j=\int ds \,[\hat{\rho}_{R j}(s)+\hat{\rho}_{L j}(s)]\enspace
\end{eqnarray}
for $j=1,2$.
 Thus, an elementary excitation with charge $q_j=1$ corresponds
to a kink of amplitude $\sqrt{\frac{\pi}{K}}$ in the bosonic field $\hat{\Phi}_j$. 
Further zero modes \cite{Eggertreview} of the fields $\hat{\Phi}_j$ and $\hat{\Theta}_j$   govern the commutation relations 
\begin{eqnarray}
[\hat{q}_j, \hat{\psi}_{\alpha j^\prime}(s)]=\delta_{j,j^\prime}\hat{\psi}_{\alpha j^\prime}(s)\enspace
\end{eqnarray}
such that acting with a left- or right-moving fermion field operator of the form of  Eq.~(\ref{eq:Bosonization_identity_rescaled}) indeed changes the total number of fermions and total charge relative to the ground state in integral numbers as physically demanded.

In this bosonized representation, the total fermionic Hamiltonian $\hat{H}=\hat{H}_{1P}+\hat{H}_\mathrm{int}$ of Eqns.~(\ref{eq:H_single_particle_fermion}) and (\ref{eq:H_Interaction_fermionic}) is
mapped onto two single-channel Luttinger liquids that are coupled
at $s=0$ by a chiral impurity scattering term:
\begin{eqnarray}
\hat{H} & = & \frac{1}{2}\sum_{j=1,2}\int ds\,\left[\left(\partial_{s}\hat{\Phi}_{j}\right)^{2}+\left(\partial_{s}\hat{\Theta}_{j}\right)^{2}\right]\nonumber \\
 &  & +\,g_{F}\,\hat{t}_{F}+g_{B}\,\hat{t}_{B} +G_{F}\,\hat{T}_{F}+G_{B}\,\hat{T}_{B}\;,\label{eq:Bosonized_rescaled}
\end{eqnarray}
where we write the single-fermion tunnelling operators at the intersection in bosonized form as
\begin{widetext}
\begin{align}
\hat{t}_{F} & =\frac{-2\sigma_{x}}{\pi\lambda}\,\cos\left[\sqrt{\pi K}\left(\hat{\Phi}_{1}(0)-\hat{\Phi}_{2}(0)\right)\right]\sin\left[\sqrt{\pi/K}\left(\hat{\Theta}_{1}(0)-\hat{\Theta}_{2}(0)\right)\right]\label{eq:single-particle-forward}\\
\hat{t}_{B} & =-\frac{2\sigma_{y}}{\pi\lambda}\,\sin\left[\sqrt{\pi K}\left(\hat{\Phi}_{1}(0)+\hat{\Phi}_{2}(0)\right)\right]\cos\left[\sqrt{\pi/K}\left(\hat{\Theta}_{1}(0)-\hat{\Theta}_{2}(0)\right)\right]\;\label{eq:single-particle-backward}
\end{align}
\end{widetext}
and the two-fermion operators as
\begin{eqnarray}
\hat{T}_{F} & = & \frac{1}{2\pi^2\lambda^2}\,\cos\left[\sqrt{4\pi K}\,\left(\hat{\Phi}_{1}(0)-\hat{\Phi}_{2}(0)\right)\right]\label{eq:two-particle-forward}\\
\hat{T}_{B} & = & -\frac{1}{2 \pi^2\lambda^2}\,\cos\left[\sqrt{4\pi K}\,\left(\hat{\Phi}_{1}(0)+\hat{\Phi}_{2}(0)\right)\right].\label{eq:two-particle_backward}
\end{eqnarray}

\subsection{Renormalization group analysis}

In an approach analogous to that of Komnik and Egger \cite{Komnik1998},
we now study the competition between single-particle 
terms like $\hat{t}_{F,B}$ and two-particle terms like $\hat{T}_{F,B}$. 
Both single-particle terms have conformal
spin one and a scaling dimension of $\triangle_{F,B}^{\left(1\right)}=\left(K+1/K\right)/2$. Hence
they are both \emph{irrelevant} when interactions are repulsive, \textit{i.e.}
for $K<1$. As was pointed out by Egger and Komnik, however, one has to be careful drawing
conclusions from the scaling dimension $\left(K+K^{-1}\right)/2\geq1$
of the single-particle terms, since they have conformal
spin one and can generate higher-order terms that are relevant.

A one-loop renormalization group analysis shows, in fact,
that each of the single-particle terms $\hat{t}_{F,B}$ generates at second-order a two-particle
term $\hat{T}_{F,B}$, with scaling dimension $\triangle_{F,B}^{\left(2\right)}=2K$, and
zero conformal spin. Both $\hat{T}_{F,B}$ are thus \emph{relevant} for $K<1/2$,
and marginal for $K=1/2$. The respective operator-product expansion
(OPE) calculation is straightforward, and similar in spirit to previous
analyses of impurity \cite{Komnik1998} and bulk \cite{Yakovenko1992} perturbations.
In contrast to the case in \cite{Yakovenko1992}, though, the Luttinger parameters $u$ and $K$
in our wires are fixed, since in our problem the perturbations $\hat{t}_{F,B}$ and $\hat{T}_{F,B}$ are localized effects at $s=0$, which cannot change the bulk properties of the whole wires. Our study also differs from previous ones in that we analyse
the RG flows of forward and backward scattering terms separately, since we expect the
respective bare coupling constants to differ greatly.

Neglecting irrelevant terms, the one-loop RG equations for the flow of the effective coupling constants $g_{F,B}$
and $G_{F,B}$ as functions of the length scale $l$ (large $l$ being long wavelength, low frequency and low energy) are
\begin{eqnarray}
\frac{dg_{F,B}}{dl} & = & \left[1-\frac{K+K^{-1}}{2}\right]\,g_{F,B}\nonumber \\
\frac{dG_{F,B}}{dl} & = & (1-2K)G_{F,B}+(K-K^{-1})g_{F,B}^{2}\;.
\label{eq:RG_equations}
\end{eqnarray}
The general solutions are
\begin{eqnarray}\label{RGsol1}
	g_{F,B}(l)&=& g_{F,B}(0) e^{-\frac{(K-1)^{2}}{2K}l}\nonumber\\
	G_{F,B}(l)&=& G_{F,B}(0) e^{(1-2K)l}\\
	&&\!\!\!\!\!\!\!\!\!\!\!\!\!\!\!\! + \frac{1-K^{2}}{K^{2}+K-1}[g_{F,B}(0)]^{2}\left(e^{(1-2K)l}-e^{-\frac{(K-1)^{2}}{K}l}\right)\;.\nonumber
	\end{eqnarray}
The coupling constants of the `backward' and `forward' processes thus
evolve independently of each other under the RG flow (\ref{eq:RG_equations}).

We see that all four terms are irrelevant in the range $1/2<K<1$,
and hence in that parameter
regime the two wires would effectively decouple completely at long wavelengths. At $K=1$,
the two-body terms\emph{ $\hat{T}_{F,B}$ }are irrelevant, while the
single-particle terms \emph{$\hat{t}_{F,B}$ }are \emph{marginal}. Our RG calculation thus consistently preserves the single-particle Hamiltonian Eqn.~(\ref{eq:H_single_particle_fermion}) in the non-interacting limit.

For \emph{strong interactions} $K\leq1/2$, however, which shall be
our focus in the remainder of the paper, we see that while the single-particle
terms \emph{$\hat{t}_{F,B}$} are \emph{irrelevant}, they generate
another set of localized terms that \emph{still} couple both wires
together at $s=0$, namely the very operators $\sim \exp{i\sqrt{4\pi K}\,\left(\Phi_{1}\pm\Phi_{2}\right)}$ that we have defined as $\hat{T}_{F,B}$. These two-fermion terms are \emph{relevant
for $K<1/2$ }and\emph{ marginal} for $K=1/2$, while the single-particle $\hat{t}_{F,B}$ terms both run to zero
at low energy. 

For repulsive interactions $K<1$, there is a length scale $l\geq l*=2/\left(K+1/K-2\right)$
beyond which we can neglect the final term in $G_{F,B}$, proportional to $\exp{-l(K-1)^{2}/(2K)}$,
in comparison with the other terms. In this range of wavelengths we can write 
\begin{eqnarray}\label{tildeG}
	G_{F,B}(l)\approx\tilde{G}_{F,B}(0)\,e^{(1-2K)l}
\end{eqnarray}
with `effective bare values' for the two-particle co-efficients
\begin{equation}
\tilde{G}_{F,B}(0)\equiv G_{F,B}(0)+\left[g_{F,B}(0)\right]^{2}\frac{1-K^{2}}{K^{2}+K-1}\label{eq:effective_g_F}
\end{equation}
that are modified by the bare values of the single-particle
processes. In other words: the low-energy behaviour of the interacting system at the intersection
is dominated by localized two-particle $\hat{T}_{F,B}$
terms instead of single-particle $\hat{t}_{F,B}$ terms, but the effective low-energy strengths of the coupling
constants of the two-particle terms are still substantially affected by the single-particle bare coupling strengths.

We can now consider the competition between the two-particle `forward' and `backward' tunnelling processes
$\hat{T}_{F}$ and $\hat{T}_{B}$. We are focusing on the range of parameter values
that arises naturally in the context of topological intersections
in quantum valley Hall edge states. We therefore assume that the bare value of the single-particle `backwards' tunnelling strength $g_{B}(0)$ is much larger than the other three bare strengths $g_{F}(0)$ and $G_{F,B}(0)$.
Inserting this initial condition into our RG solutions (\ref{RGsol1}) shows that only $G_{B}$ remains significant at long wavelengths for $K\leq1/2$, and so only the two-particle tunnelling operator $\hat{T}_{B}$ need be retained to represent the low-energy effects of the intersection. That other two-particle term $\hat{T}_{F}$ will also be relevant in this case, but because its bare value is so much lower its effective strength $G_{F}$ will remain much smaller than $G_{B}$ even at increasing wavelength.

For $K<1/2$, $G_{B}$ will in fact flow to strong coupling. 
Following Kane
\cite{Kane1992,Komnik1998} we can express the strength of coupling
indirectly in this regime, by defining an energy scale $l_{c}$
which is like $\lambda_{QCD}$ in quantum chromodynamics, in that
it is the scale $l$ at which $G_{B}(l)$ grows to order
unity. From (\ref{tildeG}) we find this scale to be 
\begin{equation}
l_{c}= \frac{\ln \tilde{G}_{B}(0)}{2K-1}\label{omegac}\;.
\end{equation}
The case $K=1/2$ exactly is seen from (\ref{tildeG}) to be a special case in which $\hat{T}_{B}$ is marginal. In a certain sense it is nonetheless typical of cases with $K$ near 1/2, because if $K$ is slightly greater than 1/2, $G_{B}$ will run towards zero only slowly, and thus still remain significant for a broad range of long (but not infinite) wavelengths. Conversely, even if $K$ approaches 1/2 from below (the strongly interacting side), (\ref{omegac}) says that the wavelength $l_{c}$ beyond which the effect of $\hat{T}_{B}$ is truly strong becomes infinite in the limit $K\to 1/2$. So once again there will be a broad range of long but not infinite wavelengths on which $\hat{T}_{B}$ has a significant but not dominating effect. For any $K$ sufficiently close to 1/2, therefore, we can consistently treat $G_{B}$ as a finite parameter by focusing on an experimentally accessible energy range of long wavelengths that are yet not too long.

We will therefore focus on this marginal case for the remainder
of this paper, because although it can be considered in the sense just described to be a typical case of
`medium-strength' interactions, it is exactly solvable by the technique
known as \emph{refermionization}.

\section{The marginal case $K=1/2$}

\subsection{Refermionization}

At $K=1/2$, the operator $\hat{T}_{B}$ is marginal
and it is possible to re-express our highly non-linear bosonic Hamiltonian (\ref{eq:Bosonized_rescaled})
as a theory of non-interacting fermions. 
We begin by introducing the following bosonic fields:
\begin{eqnarray}
\hat{\varphi}_\pm(s)&=&\frac{1}{\sqrt{2}}\left(\hat{\Phi}_1(s) \pm \hat{\Phi}_2(s) \right)\\
\hat{\theta}_\pm(s)&=&\frac{1}{\sqrt{2}}\left(\hat{\Theta}_1(s) \pm \hat{\Theta}_2(s) \right)\;.
\end{eqnarray}
These new bosonic fields have the correct commutation relations for Luttinger liquid dual phase fields, $\left[\hat{\varphi}_{\pm}(s),\hat{\theta}_{\pm}(s')\right]=-\frac{i}{2}\textrm{sgn}(s-s')$
with other commutators vanishing, but since each of the $\hat{\varphi}_{\pm},\hat{\theta}_{\pm} $ fields is a sum of fields on wire 1 and wire 2, the $s$ argument of them refers to two distinct locations in physical space. We will be able to use these peculiar non-local fields to solve the time evolution of the $K=1/2$ system very straightforwardly, but we will need to include an additional step of expressing our results in terms of local observables.

\subsubsection{Local charge densities}
The reason for defining the non-local fields $\hat{\varphi}_{\pm},\hat{\theta}_{\pm} $ appears when we express the Hamiltonian  in terms of them (including only $\hat{T}_{B}$ with renormalized coupling):
\begin{eqnarray}\label{eq:Renormalized model2}
\hat{H}  &=& \frac{1}{2}\sum_{\pm}\int ds\,\left[\left(\partial_{s}\hat{\varphi}_{\pm}\right)^{2}+\left(\partial_{s}\hat{\theta}_{\pm}\right)^{2}\right] \nonumber \\
&& -\frac{2 V}{\pi\lambda}\cos\left(\sqrt{8\pi K} \hat{\varphi}_{+}(0)\right)
\end{eqnarray}
in which $V=\frac{\tilde{G}_{B}}{4\pi\lambda}$.
We thereby discover that the $\hat{\varphi}_{-}$ fields do not appear in the intersection term at all. Only the $\hat{\varphi}_{+}$ fields are involved in tunnelling between the two wires. 

As a next step, we  \emph{refermionize} by defining new fermionic quasiparticles  \begin{eqnarray}\label{eq:Bosonization_identity_rescaled2}
\hat{\Psi}_{\alpha r}(s) & = & \frac{\eta_{\alpha r}}{\sqrt{2\pi\,\lambda}}e^{i\left(\pm_\alpha\sqrt{\pi }\,\hat{\varphi}_{r}(s)-\sqrt{\pi}\,\hat{\theta}_{r}(s)\right)}
\end{eqnarray}
 for $\alpha=R,L$ and $r=\pm$ where again the  role of the Klein factors $\eta_{\alpha r}$  which obey $\left\{ \eta_{\alpha r},\eta_{\alpha' r'}\right\} =2\delta_{\alpha \alpha'}\delta_{r r'}$ is to ensure correct anti-commutation relations among the different Fermion species. The refermionization prescription is chosen such that inserting it into Eq.~(\ref{eq:Renormalized model2}) does not generate bulk interactions among the new  quasiparticles, i.e. all prefactors in the exponent must be the same as in the analogous Eq.~(\ref{eq:Bosonization_identity_rescaled}) for non-interacting fermions (corresponding to $K=1$). In general after refermionization the new fermions (`refermions') carry fractional charge which can be quantified by considering the charge operators $\hat{q}_1$ and $\hat{q}_2$ of Eq.~(\ref{chargeoperator}).
Their commutators with the refermion creation operators yield $[\hat{q}_1, \hat{\Psi}_{\alpha r}^\dagger(s)]=\sqrt{\frac{K}{2}} \,\hat{\Psi}_{\alpha r}^\dagger(s)$ and $[\hat{q}_2, \hat{\Psi}_{\alpha r}^\dagger(s)]=\pm_r \sqrt{\frac{K}{2}} \,\hat{\Psi}_{\alpha r}^\dagger(s)$  for $r=\pm$ and $\alpha=R,L$ and, thus, these refermion quasi-particles carry absolute value of charge $\sqrt{\frac{K}{2}}$ with respect to both $\hat{q}_1$ and $\hat{q}_2$. Since electric charge is still carried microscopically by unit-charged electrons, this means that a single refermion can never be created on its own.  It can only appear in combination with other refermions or with additional charge-carrying string operators \cite{Essler2015} such that the overall operator does not leave the physical Hilbert space \cite{Rufino2013,vonDelft1998}.  In terms of the densities $\hat{n}_{\alpha \pm}=\hat{\Psi}_{\alpha \pm}^{\dagger}\hat{\Psi}_{\alpha \pm}$ with $\alpha=R,L$ the relation between the physical fermions and the refermions reads  
\begin{eqnarray}\label{refermions2}
\hat{n}_{R \pm}+\hat{n}_{L \pm} &=&  \frac{1}{\sqrt{\pi}}\partial_{s}\hat{\varphi}_\pm\\
&=&\frac{1}{\sqrt{2K}}\left(\hat{\rho}_{R 1} +\hat{\rho}_{L 1}\pm\hat{\rho}_{R 2}+\hat{\rho}_{L 2}\right) \enspace. 
\end{eqnarray}
We see that for the special point $K=1/2$, where each refermion carries charge $\sqrt{\frac{K}{2}}=1/2$, we can identify the physical charge density operators on each wire $j=1,2$
as particularly simple sums and differences of the refermion densities: 
 \begin{eqnarray}\label{density0}
	\hat{\rho}_{R 1}(s)+\hat{\rho}_{L 1}(s) &=& \frac{1}{2}\left(\hat{n}_{R +} +  \hat{n}_{L +}+ \hat{n}_{R -} + \hat{n}_{L -} \right)\nonumber\\
	\hat{\rho}_{R 2}(s)+\hat{\rho}_{L 2}(s) &=& \frac{1}{2}\left(\hat{n}_{R +} +  \hat{n}_{L +}- \hat{n}_{R -} + \hat{n}_{L -} \right)\enspace.\nonumber \\
\end{eqnarray}
We will now see that $K=1/2$ also greatly simplifies the effect of our intersection impurity.

\subsubsection{The impurity at $K=1/2$}
The intersection term in (\ref{eq:Renormalized model2}) has turned out to involve only the $+$ field for any $K$, as long as we can neglect $G_{F}$. For $K=1/2$, however, we have the further great simplification that the $\hat{\varphi}_{\alpha+}$ fields each appear in the exponent of the intersection term with the prefactor $\sqrt{4\pi}$ that one has for a product of a left-moving refermionized field operator with a right-moving one. Thus, for $K=1/2$ indeed the intersection can be expressed by refermions without additional string operators and naturally the intersection will not lead out of the physical Hilbert space. 
Choosing a representation of products of the Klein factors such that $\eta_{R+}^{\dagger}\eta_{L+}=-i$,
we can thereby cast the complete total Hamiltonian at $K=1/2$ as 
\begin{eqnarray}
\hat{H} & = &  \sum_{\alpha=R/L}\sum_{r=\pm}\int ds\,\left\{\hat{\Psi}_{\alpha r}^{\dagger}\left(\mp_\alpha i\,\partial_{s}\right)\hat{\Psi}_{\alpha r}\right\}\nonumber \\ && -i\,2\,V \,\left\{ \,\hat{\Psi}_{R+}^{\dagger}\hat{\Psi}_{L+}\left(0\right)-\hat{\Psi}_{L+}^{\dagger}\hat{\Psi}_{R+}\left(0\right)\right\}\;,\label{eq:Refermionized_Hamiltonian}
\end{eqnarray}
in which the intersection term is now merely a \emph{single-particle} scattering impurity at $s=0$. 
\begin{widetext}
We can diagonalize this $\hat{H}$ straightforwardly by expanding the refermionized field operators in the basis of single-particle energy eigenstates,
\begin{eqnarray}\label{expand}
	\hat{\Psi}_{R+}(s) &=&\frac{1}{\sqrt{2\pi}} \int\!dk\, e^{iks}\Bigl(\theta(-s)\hat{a}_{Rk+}+\theta(s)[\mathcal{T}\hat{a}_{Rk+}-\mathcal{R}\hat{a}_{Lk+}]\Bigr)\nonumber\\
	\hat{\Psi}_{L+}(s) &=&\frac{1}{\sqrt{2\pi}}\int\!dk\,e^{-iks}\Bigl(\theta(s)\hat{a}_{Lk+}+\theta(-s)[\mathcal{T}\hat{a}_{Lk+}+\mathcal{R}\hat{a}_{Rk+}]\Bigr)\nonumber\\
	\hat{\Psi}_{R-}(s) &=&\frac{1}{\sqrt{2\pi}}\int\!dk\,e^{iks}\hat{a}_{Rk-}\nonumber\\
	\hat{\Psi}_{L-}(s) &=&\frac{1}{\sqrt{2\pi}}\int\!dk\,e^{-iks}\hat{a}_{Lk-}\;,
\end{eqnarray}
where $\theta(s)$ is the Heaviside step function and the transmission and reflection co-efficients are
\begin{equation}\label{}
	\mathcal{R}=\frac{2V}{1+V^{2}}\qquad\qquad	\mathcal{T}=\frac{1-V^{2}}{1+V^{2}}\;,
\end{equation}
which obey $\mathcal{R}^{2}+\mathcal{T}^{2}=1$. 
\end{widetext}

One can check by straightforward integration that inserting the expansions (\ref{expand}) into the refermionized Hamiltonian (\ref{eq:Refermionized_Hamiltonian}) produces
\begin{eqnarray}\label{Hdiag}
	H = \sum_{\alpha=L,R}\sum_{\pm}\int\!dk\,k\, \hat{a}^{\dagger}_{\alpha k\pm}\hat{a}_{\alpha k\pm}\;.
\end{eqnarray}
We therefore see that negative $k$ modes in the expansion (\ref{expand}) have negative energy with respect to the Fermi energy, and so
the usual particle-hole transformation will mean re-defining $\hat{a}_{\alpha k\pm}\to \hat{c}^{\dagger}_{\alpha k\pm}$ for all $k<0$. Normal ordering will then mean moving all daggered operators to the left, after performing the particle-hole transformation.

We have now diagonalized the many-body Hamiltonian (\ref{eq:Refermionized_Hamiltonian}) into (\ref{Hdiag}) through the expansion (\ref{expand}) of the refermionized field operators into fermionic creation and destruction operator for orthogonal normal modes; and we have expressed the observable charge density operator in terms of bilinear functions of the refermionized field operators. We are therefore ready to compute the exact time-dependent expectation values not only of the charge density itself, but of any functions or functionals of the charge density, for our Luttinger intersection at $K=1/2$. The only remaining question is: What initial quantum states do we want to evolve?

\subsection{Incident charge density waves}
For intersecting wire problems like ours, it has been customary to compute conductances for DC currents through the intersection that may result from differing voltages (chemical potentials) applied to the wire leads. The simplicity of the Luttinger intersection problem in our $K=1/2$ case, however, will allow us to compute much more general time-dependent results for propagation of charge density wave packets through the intersection, as in the experiments reported in Ref.~[\onlinecite{Hashisaka2017}]. Continuous AC waves can be recovered in the limit of extremely broad packets. Continuous DC currents can also be represented by taking very broad packets of very long wavelength, since the region near the crest of a moving wave looks just like steady current.

States with incident charge density wave packets can be prepared in experiments by applying localized time-dependent potentials far away from the wires' intersection, for example by shining maser pulses onto the leads, or by applying time-dependent voltages via capacitively coupled gate electrodes \cite{Hashisaka2017}. Both these methods effectively apply arbitrary time- and space-dependent external potentials that couple to the local charge densities in the wires. We show in Appendix B that driving our refermionized system (\ref{eq:Refermionized_Hamiltonian}) in this way can prepare an initial quantum state with arbitrary charge density wave packets converging onto the intersection from all four leads. 

We further show in Appendix B that the evolution under  (\ref{eq:Refermionized_Hamiltonian}) of this initial state can be represented simply if we use the Heisenberg picture of quantum mechanics, in which quantum states are time-independent and quantum fields evolve in time. In the Heisenberg picture, the quantum state which represents the experimental initial state with incident charge density waves will be simply the ground state of (\ref{Hdiag}) (\textit{i.e.}, the state which is annihilated by all $\hat{a}_{\alpha k\pm}$ for $k>0$ and by all $\hat{a}^{\dagger}_{\alpha k\pm}$ for $k<0$). The initial charge density waves will be fully and exactly represented by multiplying the time-dependent Heisenberg field operators by certain c-number phases. Other than these c-number pre-factors that represent the initial charge density wave packets, the only Heisenberg evolution of the fields will be that generated by (\ref{Hdiag}), namely $\hat{a}_{\alpha k\pm}(t)=\hat{a}_{\alpha k\pm}e^{-ikt}$.

The result that we derive in Appendix B is\begin{widetext}
\begin{eqnarray}\label{CDWsol}
\hat{\Psi}_{R-}(s,t)&=& \frac{1}{\sqrt{2\pi}}e^{-i\mathcal{A}_{R-}(s-t)}\int\!dk\,e^{ik(s-t)}\hat{a}_{Rk-}\nonumber\\
\hat{\Psi}_{L-}(s,t)&=& \frac{1}{\sqrt{2\pi}}e^{-i\mathcal{A}_{L-}(s+t)}\int\!dk\,e^{-ik(s+t)}\hat{a}_{Lk-}\\
\hat{\Psi}_{R+}(s,t)&=&\frac{1}{\sqrt{2\pi}}\int\!dk\,e^{ik(s-t)}\left([\theta(-s)+\mathcal{T}\theta(s)]e^{-i\mathcal{A}_{R+}(s-t)}\hat{a}_{Rk+}
				- \mathcal{R}\theta(s) e^{-i\mathcal{A}_{L+}(t-s)}\hat{a}_{Lk+}\right)\nonumber\\ 
\hat{\Psi}_{L+}(s,t)&=&\frac{1}{\sqrt{2\pi}}\int\!dk\,e^{-ik(s+t)}\left([\theta(s)+\mathcal{T}\theta(-s)]e^{-i\mathcal{A}_{L+}(s+t)}\hat{a}_{Lk+}
				+ \mathcal{R}\theta(-s) e^{-i\mathcal{A}_{R+}(-t-s)}\hat{a}_{Rk+}\right)\;.\nonumber
\end{eqnarray}
Here the c-number functions $\mathcal{A}_{\alpha\pm}$ define the initial charge density wave packets; we will discuss some explicit examples below. From now on we will express all of our results in terms of $\mathcal{R}$ and $\mathcal{T}$ rather than of the intersection strength $V$ in (\ref{eq:Refermionized_Hamiltonian}). We will assume that a wide range of $\mathcal{R}$ and $\mathcal{T}$ values are possible, for suitable microscopic parameters of the wires and their intersection, and we will choose particular $\mathcal{R}$ and $\mathcal{T}$ values for illustrative purposes, to show most clearly the kind of qualitative behavior that can result from the Luttinger intersection.

Although it is easy to confirm that (\ref{CDWsol}) is a solution to the Heisenberg equations of motion generated by (\ref{eq:Refermionized_Hamiltonian}), it may seem surprising that the entire effect of the driving fields is to give the fields c-number phase factors. If one naively constructs the charge density (\ref{density0}), one might suppose $\hat{\Psi}_{\alpha\pm}^{\dagger}\hat{\Psi}_{\alpha\pm}$ to be completely unaffected by any c-number phase factors, and therefore conclude that we have no charge density waves at all. This conclusion would be wrong, however, because this is one of the points at which we must recall that our quantum fields must all be projected into the  subspace of many-body Hilbert space within which the fermion dispersion relation is linear, by smearing out short wavelengths. As we review in Appendix B, the actual results when this is properly done turn out to be, for example,
\begin{eqnarray}\label{}
	:\!\hat{\Psi}_{\alpha-}^{\dagger}(s,t)\hat{\Psi}_{\alpha-}(s,t)\!: =\frac{1}{2\pi}\frac{\partial}{\partial s}\mathcal{A}_{\alpha-}(s-t)\ +\  \frac{1}{\sqrt{2\pi}}\int\!dk dk'\,e^{i(k-k')s}:\!\hat{a}^{\dagger}_{\alpha k'-}\hat{a}_{\alpha k-}\!: \;,
\end{eqnarray}
where $:\dots :$ denotes normal ordering after the particle-hole transformation for all negative $k$ modes. Thus $(2\pi)^{-1}\partial_{s}\mathcal{A}_{\alpha-}$ are precisely the charge density waves that we wish to study. The more complicated effects of the $\mathcal{A}_{\alpha+}$ phase factors in the $\hat{\Psi}_{\alpha+}(s,t)$ fields will be the main results of our paper.

\subsection{Heisenberg time evolution of charge density}
We can now insert our Heisenberg evolutions (\ref{CDWsol}) into our expressions (\ref{density0}) for the local charge densities, performing the correct spatial smearing as explained in Appendix B, and discarding all constant Fermi sea contributions.
If we define $\pm_{j}$ to be $+1$ for $j=1$ and $-1$ for $j=2$, and use the identity $\mathcal{T}^{2}=1-\mathcal{R}^{2}$, then we find that the Heisenberg-picture time-dependent charge densities can be written as sums of three kinds of terms:
\begin{eqnarray}\label{rhoPsi}
\hat{\rho}_{j}(s,t) &=& \hat{\rho}^{0}_{j}(s,t)+\mathrm{sgn}(s)\left[\hat{\rho}_{C}(|s|-t)+\hat{\rho}_{X}(|s|-t)\right]\;.
\end{eqnarray}
Here $\hat{\rho}_{j}^{0}$ is simply the charge propagation that we would have if we had two separate wires, with no intersection:
\begin{eqnarray}\label{rho012}
	\hat{\rho}^{0}_{j}(s,t)&=&\int\!\frac{dk dk'}{4\pi}\,\left[e^{i(k-k')(s-t)}\left(:\!\hat{a}^{\dagger}_{Rk'+}\hat{a}_{Rk+}\!:\pm_{j}:\!\hat{a}^{\dagger}_{Rk'-}\hat{a}_{Rk-}\!:\right)+e^{-i(k-k')(s+t)}\left(:\!\hat{a}^{\dagger}_{Lk'+}\hat{a}_{Lk+}\!:\pm_{j}:\!\hat{a}^{\dagger}_{Lk'-}\hat{a}_{Lk-}\!:\right)\right]\nonumber\\
&&-\frac{1}{4\pi}\frac{\partial}{\partial s}\left[\Bigl(\mathcal{A}_{R+}(s-t)+\mathcal{A}_{L+}(s+t)\right)\pm_{j}\left(\mathcal{A}_{R-}(s-t)+\mathcal{A}_{L-}(s+t)\Bigr)\right]\;,
\end{eqnarray}
where we take $\pm_{j}\to+$ for $j=1$ and $\pm_{j}\to-$ for $j=2$. 

The other two terms $\hat{\rho}_{C,X}$ in (\ref{rhoPsi}) are the same for both wires 1 and 2, and their contributions to the charge density are both odd functions of position $s$ along the wires. We distinguish them as two separate terms because they are of significantly different forms. The first term has been labelled with a `C' subscript because it is more conventional and convenient, in that it is a combination of local charge density operators of the usual kind, with separate products of left- and right-moving operators:
\begin{eqnarray}\label{rhoC}
	\hat{\rho}_{C}(|s|,t)&=&\frac{\mathcal{R}^{2}}{4\pi}\int\!dk dk'\,e^{i(k-k')(|s|-t)}\left(:\!\hat{a}^{\dagger}_{Lk'+}\hat{a}_{Lk+}\!:-:\!\hat{a}^{\dagger}_{Rk+}\hat{a}_{Rk+}\!:\right)\nonumber\\
		&&+\frac{1}{4\pi}\frac{\partial}{\partial s}\left(\mathcal{A}_{L+}(t-|s|)-\mathcal{A}_{R+}(|s|-t)\right)\;.
\end{eqnarray}
The final term $\hat{\rho}_{X}$ is in contrast more exotic, in that it is a cross term which mixes left- and right-moving modes:
\begin{eqnarray}\label{rhoX}
	\hat{\rho}_{X}(|s|-t)&=&\frac{\mathcal{RT}}{4\pi}\int\!dk dk'\,\left(e^{i(k-k')(|s|-t)}e^{i[\mathcal{A}_{R+}(|s|-t)-\mathcal{A}_{L+}(t-|s|)]}\hat{a}^{\dagger}_{Rk'+}\hat{a}_{Lk+}+\mathrm{H.c.}\right)\;.
\end{eqnarray}

We can note that both $\hat{\rho}^{0}_{j}$ and $\hat{\rho}_{C}$ are simply sums of quantum terms and classical terms, neither of which involves the other. The more exotic term $\hat{\rho}_{X}$, however, is not a simple sum of separate quantum and classical pieces. To understand what all these terms mean, we can now proceed to compute some experimentally observable expectation values.

\subsection{Expected charge density}
All the operators in (\ref{rhoPsi}) have ground state expectation value zero; their effects are only seen in higher-order correlation functions. The time-dependent expectation value of the charge density is therefore given entirely by the classical terms in (\ref{rho012}) and (\ref{rhoC}), consisting of various sums of derivatives of $\mathcal{A}_{\alpha\pm}$ functions. To see the effects of our Luttinger intersection we will now focus on a simple scenario in which the derivatives of the  $\mathcal{A}_{\alpha\pm}$ functions are all Gaussian packets (so the $\mathcal{A}_{\alpha\pm}$ themselves are various integrals of Gaussians). Specifically, we take
\begin{eqnarray}\label{AA}
	\frac{1}{2\pi}\partial_{s}\mathcal{A}_{R\pm}&=&A_{R1}e^{-\frac{(s+D-t)^{2}}{\Lambda^{2}}}\cos[k_{0}(s+D-t)-\delta_{R1}] \pm A_{R2}e^{-\frac{(s+D-t)^{2}}{\Lambda^{2}}}\cos[k_{0}(s+D-t)-\delta_{R2}]\nonumber\\
	\frac{1}{2\pi}\partial_{s}\mathcal{A}_{L\pm}&=&A_{L1}e^{-\frac{(s-D+t)^{2}}{\Lambda^{2}}}\cos[k_{0}(s-D+t)-\delta_{L1}] \pm A_{L2}e^{-\frac{(s-D+t)^{2}}{\Lambda^{2}}}\cos[k_{0}(s-D+t)-\delta_{L2}]\;.
\end{eqnarray}
Here $D$ represents the initial distance of the packets from the origin, while $\Lambda$ is their width; $\Lambda$ is much smaller than $D$ but much larger than the wavelength $1/k_{0}$. The amplitudes $A_{\alpha j}$ and phase shifts $\delta_{\alpha j}$ represent experimentally tunable parameters. In general the four packets could all begin at different distances from the intersection, but interesting effects only appear if they overlap, and so we set all initial distances equal to $D$.

Inserting these particular $\mathcal{A}_{\alpha\pm}$ in $\langle 0|\hat{\rho}_{j}|0\rangle$, we find
 \begin{eqnarray}\label{}
\langle0|\hat{\rho}_{j}(s,t)|0\rangle &=& A_{Rj}e^{-\frac{(s+D-t)^{2}}{\Lambda^{2}}}\cos[k_{0}(s+D-t)-\delta_{Rj}]+ A_{Lj}e^{-\frac{(s-D+t)^{2}}{\Lambda^{2}}}\cos[k_{0}(s-D+t)-\delta_{Lj}] \\
 && + \mathrm{sgn}(s) \frac{\mathcal{R}^{2}}{2}e^{-\frac{(|s|+D-t)^{2}}{\Lambda^{2}}}\sum_{j'=1,2}\left(A_{Lj'}\cos[k_{0}(|s|+D-t)+\delta_{Lj}]-A_{Rj'}\cos[k_{0}(|s|+D-t)-\delta_{Rj}]\right)\;.\nonumber
\end{eqnarray}
For $t\ll D$, the whole $\mathcal{R}^{2}$ term is negligible for all $s$, and we simply have four charge density wave packets converging on the origin from all four directions, with independent amplitudes and phases. 

For $t\gg D$, however, when all four packets have propagated through the intersection, the $\mathcal{R}^{2}$ term is no longer vanishing. If we look near $s = t-D$ in this limit, far to the right at late times, we will see
\begin{eqnarray}\label{outgoingR}
\langle0|\hat{\rho}_{j}(t-D+\Delta s,t)|0\rangle &=&A_{Rj}\cos[k_{0}\Delta s-\delta_{jR}] + \frac{\mathcal{R}^{2}}{2}\sum_{j'=1,2}\left(A_{Lj'}\cos[k_{0}\Delta s+\delta_{Lj}]-A_{Rj'}\cos[k_{0}\Delta s-\delta_{Rj}]\right)
\end{eqnarray}
while if we look far to the left near $s=-(t-D)$ at late times we will see
\begin{eqnarray}\label{outgoingL}
\langle0|\hat{\rho}_{j}(D-t+\Delta s,t)|0\rangle &=& A_{Lj}\cos[k_{0}\Delta s+\delta_{jL}] - \frac{\mathcal{R}^{2}}{2}\sum_{j'=1,2}\left(A_{Lj'}\cos[k_{0}\Delta s+\delta_{Lj}]-A_{Rj'}\cos[k_{0}\Delta s-\delta_{Rj}]\right)\;.
\end{eqnarray}
\end{widetext}

A few examples will show that the intersection provides non-trivial scattering of the incident charge density waves. Consider the case of a single wave packet incident from the north-west lead; this means that $A_{1R}$ is the only non-zero amplitude. At late times we find this packet transmitted to the north-east in wire 1 with reduced amplitude $A_{1R}(1-\mathcal{R}^{2}/2)$, and reflected back to the north-west in wire 1 with amplitude $A_{1R}\mathcal{R}^{2}/2$. Transmitted waves also appear in wire 2, with equal and opposite amplitudes $\pm A_{1R}\mathcal{R}^{2}/2$ in the two directions. The total charge traveling outwards from the intersection is thus exactly equal to the incident charge. 

If we take the limit $k_{0}\to0$ in this single-incident-packet scenario, and consider the middle of the packet with $\delta_{1R}=0$ to represent DC current, then we would say that input driving power sufficient to create a current $I\propto A_{1R}$ in a wire with no intersection is able to pass a reduced current $(1-\mathcal{R}^{2}/2)I$ through wire 1 past the intersection, while exciting a parallel current $\mathcal{R}^{2}I/2$ in wire 2 through Coulomb dragging at the intersection. The total current from left to right through the two-wire system is the same as the driving source would have induced in a single wire, but the intersection spreads some of the current from wire 1 to wire 2. 

Now look at cases with two incident packets; see Fig.~3. If the two packets are both incident in wire 1, coming from both left and right with equal amplitudes but arbitrary phases, then the packets excited in wire 2 by Coulomb dragging at the intersection will have amplitudes proportional to $\mathcal{R}^{2}\sin[(\delta_{L1}+\delta_{L2})/2]$. If on the other hand our two incident packets are in separate wires, consider them both to come from the left with equal amplitude, by setting $A_{2R}=A_{1R}=A$ and $A_{1L}=A_{2L}=0$. From (\ref{outgoingR}) we find a more complicated phase-dependent transmission pattern. For $\delta_{2R}=\delta_{1R}$ it reduces to two identical packets transmitted to the right in both wires, but with reduced amplitude $(1-\mathcal{R}^{2})A$, while two identical reflected packets propagate to the left with amplitude $\mathcal{R}^{2}A$. If the two drives and packets are exactly out of phase, however ($\delta_{2R}=\delta_{1R}+\pi$), then both packets are transmitted in their wires without any reflection or attenuation, but only a phase shift. In effect the two-wire system in this case can be said to have an impedance which depends on the relative phase of the incident packets. Both of these phase dependence effects in two-packet transmission may potentially be exploited for interferometry.
\begin{figure}[H]
\center\includegraphics[width=0.9\columnwidth]{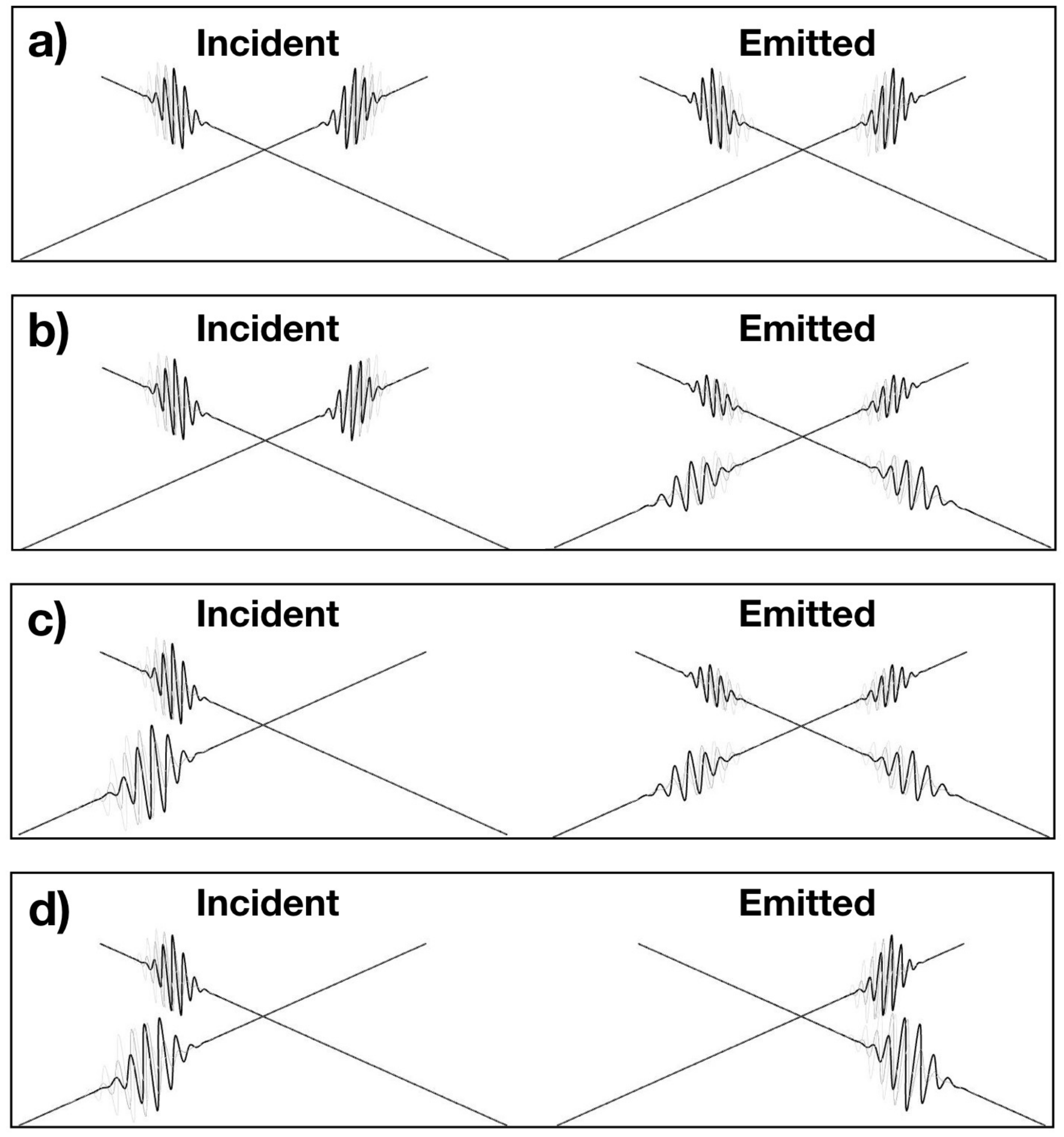}\caption{\label{fig:2Packets}Expectation value of the local charge density along the two intersecting wires, for various pairs of incident charge density wave packets, all in the illustrative case $\mathcal{R}^{2}=\mathcal{T}^{2}=1/2$. The height of the ripples is the local charge density. All packets are identical except for wave phase; in particular their patterns of charge density are either equal or opposite in sign. The two packets in each pair are timed to overlap at the intersection. Plots labeled `Incident' show the charge density at a time shortly before the packets have reached the intersection; plots labeled `Emitted' show the outgoing packets shortly after the packets have traversed the intersection. In a) and b) incident packets come from opposite ends of wire 1, with phases equal (a) or opposite (b). In c) and d) the incident packets come from the left in both wires, with phases equal (c) or opposite (d). The relative phase of the two packets determines whether or not they will split as they pass through the intersection.}
\end{figure}

A particularly striking example appears if we have four incident packets with equal amplitude $A_{\alpha j}=A$, but set the phases $\delta_{Rj}=\delta$, $\delta_{Lj}=\pi-\delta$. In this case the outgoing packets in all four leads have the same reduced amplitude $A(1-2\mathcal{R}^{2})$. If the case $\mathcal{R}^{2}=1/2$ could be achieved, the incident packets would all annihilate each other, with no outgoing packets surviving! See Fig.~4. Even for less extreme values of $\mathcal{R}$, however, it is clear that the intersection is somehow reducing the total intensity of the incident waves.

In fact this reduction occurs in general, even with fewer incident packets. This may be surprising, because we have seen that $\mathcal{R}^{2}+\mathcal{T}^{2}=1$, implying exact conservation of refermions through the intersection. Our waves here are in the expectation value of the charge density, however, rather than of refermions, and we must note that it is indeed the squared amplitude $\mathcal{R}^{2}$ which appears in (\ref{outgoingR}) and (\ref{outgoingL}), rather than $\mathcal{R}$ and $\mathcal{T}$ themselves. This means that total charge emitted from the intersection, which is proportional to the amplitude of the charge density waves, is always exactly equal to the total incident charge. 

\begin{figure}
\center\includegraphics[width=0.9\columnwidth]{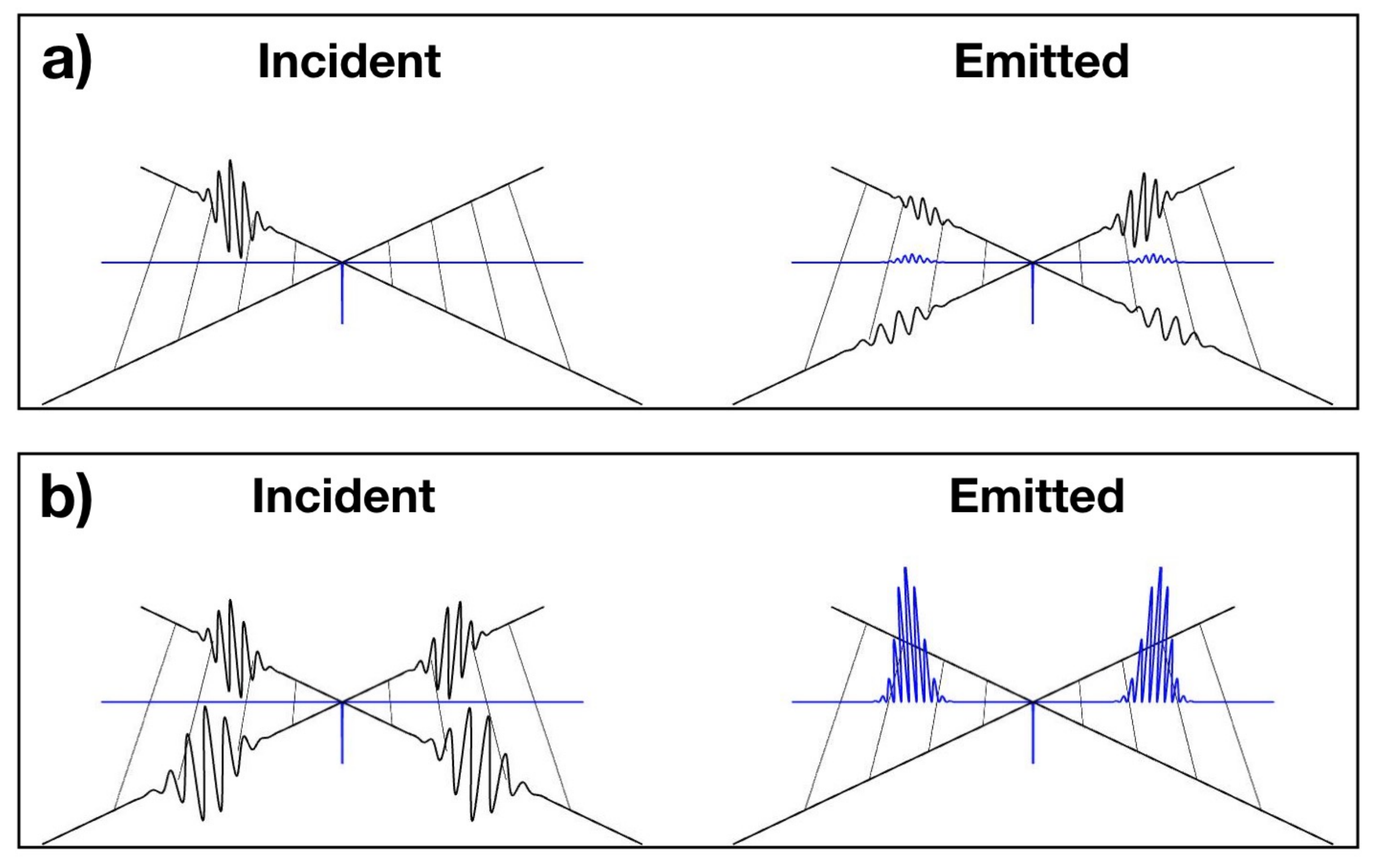}\caption{\label{Decoherence} Diagonally crossing axes show the expectation value of the charge density as in Fig.~3, again for the illustrative case $\mathcal{R}^{2}=\mathcal{T}^{2}=1/2$. Panel a) shows a single incident wave packet dividing at the intersection into four outgoing packets. Panel b) shows four converging packets, identical except that the packets coming from the right have opposite phase; the packets of charge density expectation value annihilate each other at the intersection. In both panels the horizontal blue axis shows the density-density correlation function between the two wires, at the points joined by the vertical lines. For the single incident packet in a), some of the incident energy and information is emitted in correlated quantum noise that is generated as the packet crosses the intersection. For the four incident packets in b), the entire incoming signal is transformed into quantum noise with zero expectation value but non-zero two-point correlation function.}
\end{figure}
The intensities of charge density waves are proportional to the squares of their amplitudes, however, so terms with $\mathcal{R}^{4}$ will appear for the wave intensities in our charge density expectation values, and the total ratio of emitted intensity to incident need not sum to one. Specifically, in fact, Eqn.~(\ref{outgoingR}) and (\ref{outgoingL}) imply that the total integrated intensity of incident waves in the expectation value of charge density is proportional to $A_{L1}^{2}+A_{L2}^{2}+A_{R1}^{2}+A_{R2}^{2}$, while the total emitted intensity in the expectation values is proportional to
\begin{eqnarray}\label{}
	&&A_{L1}^{2}+A_{L2}^{2}+A_{R1}^{2}+A_{R2}^{2}\nonumber\\
&&-\mathcal{R}^{2}\mathcal{T}^{2}\left(A_{R1}+A_{R2}-A_{L1}-A_{L2}\right)^{2}\;,
\end{eqnarray}
which in general is less. 

This does not of course mean that the intersection is destroying energy or information. All the waves that we have so far discussed are patterns in the expectation value of the charge density, which is simply the average of the charge densities that are observed in many runs of the same experiment. What our result therefore means is that, unless $\mathcal{RT}$ or $A_{R1}+A_{R2}-A_{L1}-A_{L2}$ should happen to vanish, the intersection transfers some of the incident information and energy from classical charge density waves, which look the same in every run of the experiment, into quantum fluctuations which vary randomly from run to run, and over many runs average to zero. 

These fluctuations whose average value is zero can systematically carry information and energy, however, because they are correlated. To see this, we can compute the time-dependent density-density correlation functions.

\begin{widetext}
\subsection{Charge density correlation functions}
The density-density correlation function in the ground state is defined as
\begin{eqnarray}\label{Sij}
	S_{ij}(s,s',t)&=&\langle0|\hat{\rho}_{i}(s,t)\hat{\rho}_{j}(s',t)|0\rangle - \langle0|\hat{\rho}_{i}(s,t)|0\rangle\,\langle0|\hat{\rho}_{j}(s',t)|0\rangle\;.
\end{eqnarray}
Inserting (\ref{rhoPsi}) into (\ref{Sij}) yields
\begin{eqnarray}\label{Sij2}
	S_{ij}(s,s',t)&=&-\frac{\delta_{ij}}{4\pi^{2} (s-s')^{2}} -\frac{\mathcal{R}^{2}}{8\pi^{2}}\frac{\mathrm{sgn}(s)\,\mathrm{sgn}(s')}{(|s|+|s'|)^{2}}+S_{X}(s,s',t)
\end{eqnarray}
where
\begin{eqnarray}\label{SX}
S_{X}(s,s',t)&=&\frac{\mathcal{R}^{2}\mathcal{T}^{2}}{4\pi^{2}}\,\mathrm{sgn}(s)\,\mathrm{sgn}(s')\,\frac{\sin^{2}\left(\frac{\mathcal{A}_{R+}(|s|-t)-\mathcal{A}_{R+}(|s'|-t)+\mathcal{A}_{L+}(t-|s'|)-\mathcal{A}_{L+}(t-|s|)}{2}\right)}{(|s|-|s'|)^{2}}
\end{eqnarray}
is the same whichever wires the points $s$ and $s'$ may be on.

The first term in (\ref{Sij2}) is simply the usual Luttinger liquid correlation function; it is due to ground state fluctuations and is present regardless of any experimentally generated charge density waves. It describes correlations decaying quadratically with distance between any two points in each wire, but not between any points on different wires. The second term in (\ref{Sij2}) is also a time-independent property of our system's ground state; it is a sort of vacuum polarization effect localized around the $s$ origin. It implies that the charge densities on the two wires become correlated near the intersection.

The final term $S_{X}$ is only non-zero if there are incident wave packets. With our incident packets defined by (\ref{AA}), $\mathcal{A}_{R+}(x)$ is a constant independent of $x$ for $x<-D$, and $\mathcal{A}_{L+}(x)$ is likewise constant for $x<D$. Hence for all $t<D$, the $S_{X}$ vanishes, and the correlation function is simply equal to its time-independent ground state value without any wave packets. This tells us that until the incident charge density wave packets reach the intersection, they are really classical waves just like laser pulses, with no effect on any quantum fluctuations. If the incident packets are followed in many successive experimental runs, they will appear exactly the same in every experiment, with only the same quantum noise superposed that is observable in the wires in their ground state, without any incident waves. 

Once $t>D$ and our packets have reached the intersection, however, $S_{X}$ becomes non-zero. For long-wavelength wave packets, the approximation
\begin{eqnarray}\label{}
	\mathcal{A}_{R+}(|s|-t)\doteq \mathcal{A}_{R+}(\frac{|s|+|s'|}{2}-t) + \mathcal{A}'_{R+}(\frac{|s|+|s'|}{2}-t)\frac{|s|-|s'|}{2}
\end{eqnarray}
remains excellent until $|s|-|s|'$ is so large that the $(|s|-|s'|)^{2}$ denominator in (\ref{SX}) makes $S_{X}$ negligible anyway. We can therefore approximate
\begin{eqnarray}\label{SX1}
	S_{X}(s,s',t)&\doteq&\frac{\mathcal{R}^{2}\mathcal{T}^{2}}{16\pi^{2}}\,\mathrm{sgn}(s)\,\mathrm{sgn}(s')\,
\left[\frac{\partial}{\partial s}\left(\mathcal{A}_{R+}(\frac{|s|+|s'|}{2}-t)-\mathcal{A}_{L+}(t-\frac{|s|+|s'|}{2})\right)\right]^{2}\nonumber\\
&\to&\frac{\mathcal{R}^{2}\mathcal{T}^{2}}{4}\,\mathrm{sgn}(s)\,\mathrm{sgn}(s')\,
e^{-2\frac{(\frac{|s|+|s'|}{2}+D-t)^{2}}{\Lambda^{2}}}\\
&&\qquad\times\left[\sum_{j'=1,2}\left(A_{Lj'}\cos[k_{0}(\frac{|s|+|s'|}{2}+D-t)+\delta_{Lj}]-A_{Rj'}\cos[k_{0}(\frac{|s|+|s'|}{2}+D-t)-\delta_{Rj}]\right)\right]^{2}\;, \nonumber
\end{eqnarray}\end{widetext}
where in the last line we have inserted our particular example of Gaussian wave packets. At $|s|=|s'|$ the above result becomes exact even for short-wavelength packets. 

If we look at (\ref{SX1}) for $s'=s$ on the same wire, therefore, we see that there are packets of correlated noise that exactly match the `missing' wave intensity in the average charge density that we noted in the preceding Subsection. Over many runs of the experiment one will see random run-to-run variations in the charge density that are equally likely to be positive or negative, but which are distinctly larger in both directions within the outgoing packet. Even in the extreme four-packet scenario mentioned above, where the expectation value of the outgoing packets vanishes, the packets that are invisible in the average could still be followed as propagating packets of enhanced fluctuations. These packets of charge density perturbations, positive and negative, are different in every run, and show no steady pattern in any one wire, but the apparently random charge density patterns in the two wires, and at opposite positions in each wire, maintain a consistent relationship.

The correlations between fluctuations do not only exist at $s=s'$, moreover; they extend over a packet-sized range of nearby $s$ and $s'$, within outward-moving envelopes that follow the outgoing wave packets of average charge density. The same correlations that exist between nearby $s$ and $s'$ within the outgoing packet envelopes also exist for $s'$ close to $-s$, even though at late times these points will be far apart. It makes no difference for the correlation pattern whether the two points $s$ and $s'$ are even on the same wire or not. The intersection thus induces long-ranged quantum correlations between outgoing charge density waves, even when the incident charge density waves are classical signals with no quantum correlations.

\section{Entanglement}

The many-body quantum states which exist after our charge density wave packets have crossed the intersection have long-range quantum correlations. Do they in fact show entanglement? We can address this question straightforwardly, in a way that also sheds some general light on the nature of these states, by examining a particular subspace of the many-body Hilbert space, consisting of the second-quantized fermionic excitations of two particular single-particle modes. Two fermionic modes define a four-dimensional Hilbert space which can be identified with a two-qubit Hilbert space (under certain conditions \cite{FermEntRef1,FermEntRef2,FermEntRef3,FermEntRef4} which we will satisfy). The Peres-Horodecki criterion \cite{Peres,Horodecki} then determines unambiguously whether these two qubits are entangled.

We first identify our two fermionic modes, by defining their fermionic destruction operators:
\begin{eqnarray}\label{b12def}
	\hat{b}_{1}&=&\frac{1}{\pi^{1/4}\sqrt{a}}\int\!ds'\,e^{-\frac{1}{2}(s'+D)^{2}/a^{2}}\hat{\psi}_{L+}(s',t)\nonumber\\
	\hat{b}_{2}&=&\frac{1}{\pi^{1/4}\sqrt{a}}\int\!ds'\,e^{-\frac{1}{2}(s'-D)^{2}/a^{2}}\hat{\psi}_{R+}(s',t)\;.
\end{eqnarray}
That is, $\hat{b}_{1}$ destroys a left-moving refermion somewhere within the short distance $a$ of the point $s=-D$, well to the left of the intersection, while $\hat{b}_{2}$ destroys a right-moving refermion somewhere well to the left of the intersection, near the point $s=+D$. 

Since both of these fermionic modes involve $\hat{\psi}_{\alpha+}$ fields, neither of them is localized on either one of our two wires, but rather both modes are delocalized between the two wires, as even superpositions of being on both. Since one of our modes is well to the left of the intersection while the other is well to the right, however, these two modes are definitely well separated in space. Although it might be difficult to probe these non-local modes experimentally we can still analyze them theoretically to reveal long-range entanglement in our system. 

Note well that $\hat{b}^{\dagger}_{1}\hat{b}$ is \emph{not} simply a Gaussian-weighted integral of the charge density around $s=-D$, not even for any combination of wires. That integrated charge density would be the operator we would obtain if we integrated the product $\hat{\psi}^{\dagger}_{L+}\hat{\psi}_{L+}$ of refermion field operators over $s$ around $-D$. The product of the two integrals is not the same as the integral of the product. The integrated charge density will have many eigenvalues, and they can be quite large, since there could be many excess charges within $a$ of $s=-D$; but the eigenvalues of $\hat{b}^{\dagger}_{1}\hat{b}$ are only 0 and 1. We may say that $\hat{b}^{\dagger}_{1}\hat{b}$ does not ask the question, ``How much charge is near $-D$?'' but rather, ``Is there a refermion occupying this Gaussian orbital?" There might be many more refermions near $-D$ with wave functions orthogonal to that Gaussian, but for that particular Gaussian wave function, Pauli allows no more than one refermion occupant.

For simplicity we will assume that $a$ is long enough for our linear dispersion relation to be valid on its scale, but yet very short compared to the wavelengths of our charge density wave packets as defined by the c-number functions $\mathcal{A}_{\alpha+}(s)$. This means for example that we will be able to approximate
\begin{eqnarray}\label{}
	e^{-i\mathcal{A}_{L+}(t-s')}\doteq e^{-i\mathcal{A}_{L+}(t-D)}e^{+i\mathcal{A}_{L+}'(t-D)\,(s'-D)}
\end{eqnarray}
within the range of $s'$ in which $\exp[-(s'-D)^{2}/(2a^{2})]$ has significant support. Since we will allow our charge density wave packet to have arbitrarily large amplitude even though its wavelength must be long compared to $a$, we will \emph{not} be allowed in general to further Taylor-expand the above exponentials.

With the above approximation we can use Eqn.~(\ref{CDWsol}) to express $\hat{b}_{1,2}$ directly in terms of our refermion normal mode operators $\hat{a}_{\alpha k+}$, since we can easily perform the Gaussian $s'$ integrals if we assume that the Gaussian factors in (\ref{b12def}) have negligible support for $s>0$ in the case of $\hat{b}_{1}$ and for $s<0$ in the case of $\hat{b}_{2}$. The results are simply Gaussian integrals in $k$-space:\begin{widetext}
\begin{eqnarray}\label{b12simp}
\hat{b}_{1}&=&\frac{\sqrt{a}}{\pi^{1/4}}\int\!dk\,e^{ik(D-t)}\Bigg[T e^{-i\mathcal{A}_{L+}(t-D)}e^{-\frac{a^{2}}{2}[k+\mathcal{A}_{L+}'(t-D)]^{2}}\hat{a}_{Lk+}+Re^{-i\mathcal{A}_{R+}(D-t)}e^{-\frac{a^{2}}{2}[k-\mathcal{A}_{R+}'(D-t)]^{2}}\hat{a}_{Rk+}\Bigg]\nonumber\\
\hat{b}_{2}&=&\frac{\sqrt{a}}{\pi^{1/4}}\int\!dk\,e^{ik(D-t)}\Bigg[T e^{-i\mathcal{A}_{R+}(D-t)}e^{-\frac{a^{2}}{2}[k-\mathcal{A}_{R+}'(D-t)]^{2}}\hat{a}_{Rk+}-Re^{-i\mathcal{A}_{L+}(t-D)}e^{-\frac{a^{2}}{2}[k+\mathcal{A}_{L+}'(t-D)]^{2}}\hat{a}_{Lk+}\Bigg]\;.
\end{eqnarray}

We can gain some understanding of what our many-body quantum state implies for the state of these two fermionic modes, by computing the expectation values of their occupation numbers. We can do this for any time $t$ by working in the Heisenberg picture, since (\ref{b12simp}) correctly expresses the Heisenberg time-dependence of the $\hat{b}_{1,2}$ as inherited from the Heisenberg-picture $\hat{\psi}_{\alpha+}(s,t)$ that we gave in (\ref{CDWsol}). Straightforward integrals reveal
\begin{eqnarray}\label{}
	\langle\Psi|\hat{b}^{\dagger}_{1}\hat{b}_{1}|\Psi\rangle&=&\frac{1}{2}\Big[1+T^{2}\mathrm{erf}\Big(a \mathcal{A}_{L+}'(t-D)\Big)-R^{2}\mathrm{erf}\Big(a \mathcal{A}_{R+}'(D-t)\Big)\Big]\nonumber\\
\langle\Psi|\hat{b}^{\dagger}_{2}\hat{b}_{2}|\Psi\rangle&=&\frac{1}{2}\Big[1+R^{2}\mathrm{erf}\Big(a \mathcal{A}_{L+}'(t-D)\Big)-T^{2}\mathrm{erf}\Big(a \mathcal{A}_{R+}'(D-t)\Big)\Big]\;,
\end{eqnarray}\end{widetext}
where $\mathrm{erf}(x)$ is the \emph{error function}
\begin{eqnarray}\label{}
	\mathrm{erf}(x)=\frac{2}{\sqrt{\pi}}\int_{0}^{x}\!dy\,e^{-y^{2}}\;.
\end{eqnarray}

Since $\mathcal{A}'_{\alpha+}(s)$ defines the classical charge density perturbation of our incident charge density waves, we can interpret $Q_{L}=a\mathcal{A}_{L+}'(t-D)$ and $Q_{R}=a\mathcal{A}_{L+}'(D-t)$ as the classical total charges, from the two left-moving and right-moving charge density waves respectively, within a Gaussian weighting envelope of width $a$. Both these classical charges can be positive or negative, representing charge density perturbations around the ground-state Fermi sea. These classical charge values are not directly equal to the occupation numbers of either of our refermion modes, since many orthogonal refermion modes may contribute to the total charges $Q_{L,R}$, but $Q_{L}$ and $Q_{R}$ do provide statistical biases to the occupation probabilities of our particular Gaussian modes. At small $x$ we have $\mathrm{erf}(x)=x+\mathcal{O}(x^{3})$, and so if $Q_{L,R}$ is small then the average occupation numbers for our two modes are small positive or negative perturbations around the ground state value of 1/2. (One-half is the average occupation of local fermion modes that is implied by a Fermi sea filling half of $k$-space). Since the limits of $\mathrm{erf}(x)$ as $x\to\pm\infty$ are $\pm1$, large $Q_{L,R}$ can in principle bias the occupation probabilities for our local refermion modes so strongly as to make the average occupation numbers approach zero or one. In fact $\mathrm{erf}(Q_{L,R})$ will approach quite close to 0 or 1 as soon as $|Q_{L,R}|$ rises much above 1. 

Having defined our two fermionic modes, we can now proceed to identify the projection of the pure many-body quantum state into their two-qubit subspace.
Using the two-qubit tensor product $|mn\rangle \equiv |m\rangle|n\rangle$ for $m,n = 0,1$, the density operator $\hat{\rho}$ for the two-mode subspace has matrix elements
\begin{eqnarray}\label{}
	\rho_{mn,m'n'}=\langle\Psi|m'n'\rangle\langle mn|\Psi\rangle
\end{eqnarray}
where $|\Psi\rangle$ is the many-body quantum state of our two-wire Luttinger system, which in our case is a pure state. The crucial step which makes this density matrix easy to compute explicitly is to recognize that we can express the density matrix in terms of expectation values of combinations of $\hat{b}_{1,2}$ and their conjugates, because if we define
\begin{eqnarray}\label{}
	|11\rangle = \hat{b}_{2}^{\dagger}\hat{b}_{1}^{\dagger}|00\rangle\qquad
|10\rangle = \hat{b}_{1}^{\dagger}|00\rangle\qquad
|01\rangle = \hat{b}_{1}^{\dagger}|00\rangle
\end{eqnarray}
then we have
\begin{eqnarray}\label{twoops}
	|11\rangle\langle00|\equiv\hat{b}_{2}^{\dagger}\hat{b}_{1}^{\dagger}\qquad
	|01\rangle\langle10|\equiv\hat{b}^{\dagger}_{2}\hat{b}_{1}
\end{eqnarray}
as well as their Hermitian conjugates. Hence we have, for example,
\begin{eqnarray}\label{b2b1}
	\rho_{01,10}\equiv \langle\Psi|01\rangle\langle10|\Psi\rangle = \langle\Psi|\hat{b}^{\dagger}_{2}\hat{b}_{1}|\Psi\rangle\;,
\end{eqnarray}
which we can compute because we know our many-body quantum state $|\Psi\rangle$ in terms of occupation number eigenstates of the $k$ modes, and we know $\hat{b}_{1,2}$ in terms of the $\hat{a}_{\alpha k+}$ operators. We can even use our Heisenberg evolution of $\hat{\psi}_{\alpha+}(s,t)$, in (\ref{CDWsol}), to obtain the time-dependent $\hat{b}_{1,2}(t)$ in Heisenberg picture, and thereby compute the two-qubit density matrix $\rho_{mn,m'n'}(t)$ at any time.

Products of three $\hat{b}_{1,2}$ and $\hat{b}_{1,2}^{\dagger}$ operators provide eight other off-diagonal mappings between $|mn\rangle$ and $|m'n'\rangle\not=|mn\rangle$, making twelve such off-diagonal operators in total. The diagonal projection operators can be be realized as products of two of the operators shown in (\ref{twoops}), for example
\begin{eqnarray}\label{}
	|00\rangle\langle00| &\equiv& |00\rangle\langle11|11\rangle\langle00|=\hat{b}_{1}\hat{b}_{2}\hat{b}_{2}^{\dagger}\hat{b}_{1}^{\dagger}\;.
\end{eqnarray}

A charge density wave of finite wavelength does not inject or remove net charge in a quantum wire, but only redistributes the charges present in the ground state. Our particular many-body quantum state $|\Psi\rangle$ is thus an eigenstate of total refermion number, and therefore the only operator combinations with non-vanishing expectation values have equal number of refermion creation and destruction operators. This implies that our two-qubit density matrix can only have two non-zero off-diagonal elements, namely $\rho_{01,10}$ and its complex conjugate $\rho_{10,01}$, given as expectation values in (\ref{b2b1}) above. The only other non-zero elements are the four diagonal ones
\begin{eqnarray}\label{}
	\rho_{11,11}&=&\langle\Psi|\hat{b}^{\dagger}_{1}\hat{b}_{1}\hat{b}_{2}^{\dagger}\hat{b}_{2}|\Psi\rangle\nonumber\\
	\rho_{01,01}&=&\langle\Psi|\hat{b}_{1}\hat{b}_{1}^{\dagger}\hat{b}_{2}^{\dagger}\hat{b}_{2}|\Psi\rangle\nonumber\\
	\rho_{10,10}&=&\langle\Psi|\hat{b}_{1}^{\dagger}\hat{b}_{1}\hat{b}_{2}\hat{b}_{2}^{\dagger}|\Psi\rangle\nonumber\\
	\rho_{00,00}&=&\langle\Psi|\hat{b}_{1}\hat{b}_{1}^{\dagger}\hat{b}_{2}\hat{b}_{2}^{\dagger}|\Psi\rangle\;,
\end{eqnarray}
which sum identically to one because of the canonical anti-commutation relation of the fermionic operators.

The Peres-Horodecki criterion tells us that a four-by-four density matrix implies a state which is not separable in the tensor product basis of the two qubits, if and only if the \emph{partial transpose} of the density matrix has one or more negative eigenvalues. The partial transpose of $\rho_{mn,m'n'}$ is $\tilde{\rho}_{mn,m'n'}=\rho_{mn',m'n}$. In our case, therefore, the diagonal elements of $\hat{\tilde{\rho}}$ are the same as those of $\hat{\rho}$, while the only non-zero off-diagonal elements of $\hat{\tilde\rho}$ are $\tilde\rho_{11,00}=\tilde{\rho}_{00,11}^{*}=\langle\Psi|\hat{b}^{\dagger}_{2}\hat{b}_{1}|\Psi\rangle$.

Since $\tilde{\rho}_{mn,m'n'}$ is thus block-diagonal we can easily compute its four eigenvalues by solving only quadratic equations. Even without evaluating the various expectation values it is straightforward to show that three of the four eigenvalues must be positive; the fourth one will be negative, making the state entangled, if and only if
\begin{eqnarray}\label{}
	\Big|\langle\Psi|\hat{b}^{\dagger}_{2}\hat{b}_{1}|\Psi\rangle\Big|^{2}>\langle\Psi|\hat{b}^{\dagger}_{1}\hat{b}_{1}\hat{b}_{2}^{\dagger}\hat{b}_{2}|\Psi\rangle\langle\Psi|\hat{b}_{1}\hat{b}_{1}^{\dagger}\hat{b}_{2}\hat{b}_{2}^{\dagger}|\Psi\rangle\;.
\end{eqnarray}

Inserting our expressions for $\hat{b}_{1,2}$ in terms of $\hat{a}_{\alpha k+}$, we find that this reduces to the condition
\begin{align}\label{}
	&4R^{2}T^{2}\big[\mathrm{erf}(Q_{L})+\mathrm{erf}(Q_{R})\big]^{2}\nonumber\\
&>\Big(1-\big[\mathrm{erf}(Q_{L})\big]^{2}\Big)\Big(1-\big[\mathrm{erf}(Q_{R})\big]^{2}\Big)\;.
\end{align}
Since the right-hand side of this inequality approaches zero whenever either of $|Q_{L,R}|$ is significantly greater than 1, while the left-hand side will not vanish unless $RT$ is zero or $Q_{L}=-Q_{R}$, it is perfectly possible to satisfy this inequality with moderately strong charge density waves. Our intersection can indeed generate long-range entanglement.

\section{Conclusions}

In addition to mixing and scattering incident charge density wave packets, with a dependence on relative phase that one expects for a coherent beam splitter, the intersection also transfers some of the coherent incident excitations into correlated quantum noise that propagates outwards along with the scattered classical packets. In extreme cases like the one of four incident packets at $\mathcal{R}^{2}=1/2$ in which no outgoing signals can be detected in charge density averages, the packets of correlated noise may even be the only outgoing signal.

This occurs because, although the intersection at $K=1/2$ is a linear beam splitter for refermions, it is still a highly nonlinear beam splitter for charge density waves. In terms of the bosonic fields whose quanta are Luttinger plasmons, the intersection term $\hat{T}_{B}$ is a sine function whose Taylor expansion includes arbitrarily high powers of creation and destruction operators. If we were to treat $\hat{T}_{B}$ as a local perturbation to the free and dispersionless bulk plasmons, we would see that it could annihilate many low-frequency plasmons and replace them with a single high-frequency one, or \textit{vice versa}. This means that if a quasi-classical Glauber coherent state of plasmons is affected by the intersection, it ceases to be a quasi-classical coherent state. If a coherent wave packet is split by the intersection into multiple packets, the nonlinearity of the plasmon beam splitter induces quantum entanglement between the outgoing packets. This phenomenon can be detected experimentally as correlations between local charge densities at spatially distant locations.

As explained in Appendix B, the multiplication of the fermion fields by c-number phases in the Heisenberg picture is exactly the fermionic representation of a Glauber coherent state of Luttinger plasmons. In the special case $K=1/2$ the refermionized representation lets us solve the intersection problem exactly, instead of perturbatively. We were thereby able to confirm the decoherence of incident coherent states and the generation of long-range quantum correlations in the charge density. The qualitative conclusion that this kind of thing will occur is more general.

Decoherence of incident charge density waves does not even require the fermions to be interacting. It is straightforward to repeat our calculations without any Coulomb interactions, by computing the evolution of charge density wave packets under our non-interacting Hamiltonian $\hat{H}_{1P}$ of Eqn.~(\ref{eq:H_single_particle_fermion}). The result is somewhat more complicated than in the $K=1/2$ case, because without interactions the left- and right-moving fermions are each affected by the intersection independently, and there is no mode like the `+' mode at $K=1/2$ that is unaffected by the intersection. In the end one obtains very similar expressions, however. The bosonization mapping from fermions to plasmons is valid regardless of whether the fermions have two-body interactions or not; classical charge density waves are always Glauber states of plasmons; and the intersection is always a nonlinear beam splitter for plasmons, even in the absence of inter-fermion interactions.

Except in special cases, however, there is no reason to expect the decoherence of incident coherent waves to be entirely destructive. If $\mathcal{R}$ is not too large, which should always be preventable by weakening the contact between the two wires, the reduction in total signal strength of the average charge density will remain modest. Transmission through the intersection by two overlapping incident waves will then depend on the relative phases of those incident waves, as in an optical beam splitter. The possibility of developing Luttinger interferometry should therefore be further investigated. The fact that decoherence can also depend on relative phases of the incident packets may even give Luttinger interferometry an additional read-out channel for phase information.

Decoherence and long-range many-body entanglement induced by the edge-state intersection are also interesting phenomena in their own right. Luttinger liquid theory shows that interacting fermions in one-dimensional channels behave generically at long wavelengths as non-interacting bosons, and although this mapping itself is a remarkable feature of quantum many-body dynamics, it makes it difficult in general to directly see quantum many-body effects in bulk in one dimensional systems. As in the Kondo effect, one looks to impurities for dramatic fingerprints of quantum dynamics. Here we have shown that the impurity representing an intersection between edge state modes can be a controlled source of interesting quantum correlations that may be accessible to direct observation by detecting correlations in charge density quantum noise.  These possibilities will also deserve further study.

One may ask, for example, whether the decoherence that is induced by the intersection is irreversible. In principle, it is not: our Hamiltonian is Hermitian and has a time-reversal symmetry. For each of our decohered states of quantum-entangled outgoing waves, therefore, there exists a time-reversed state of entangled incoming waves, which emerge from the intersection as purely classical signals. Experimental preparation of such entangled initial states will surely be much harder, however, than simply applying time-dependent classical voltages to the leads, as sufficed to prepare the incident classical waves. 

An interferometer requires a sequence of two beam splitters, though. If our entangled outgoing packets propagate through curving leads that bend around and meet each other a second time, then we will have correlated incident packets on the second intersection. Will some of the information which these packets are carrying as quantum noise correlations be returned, by the nonlinear action of the second intersection, into classical form? We intend to examine this question in future work.

\begin{acknowledgments}
	We are thankful for very useful discussions with Sebastian Eggert. 
	This work was supported by the Deutsche Forschungsgemeinschaft (DFG) via the research
	center SFB/TR185.
\end{acknowledgments} 

\setcounter{equation}{0} 
\setcounter{figure}{0}  
\renewcommand{\theequation}{A-\arabic{equation}}  
\renewcommand{\thefigure}{A-\arabic{figure}}

\appendix

\section{Renormalization Group Flow}

\subsection{Renormalization group equations}

Let us regard the Lagrangian corresponding to a Hamiltonian of the
form 
\begin{align*}
S & =\frac{1}{2}\int d\tau\,\int dx\,\Phi\left(\partial_{x}^{2}+\frac{1}{u^{2}}\partial_{\tau}^{2}\right)\Phi+\mbox{\ensuremath{\sum}}_{i}g_{i}V_{i},\\
 & =S_{0}+S_{I}
\end{align*}
 (where we regard only a single channel model for brevity and perform
a Wick rotation to imaginary time $\tau=it$). The first term is manifestly
invariant under a rescaling $\left(x,\tau\right)\rightarrow\left(\lambda x,\lambda\tau\right)$,
since the respective rescaling factors of the integrals and the differential
operators cancel. The second term $S_{I}$, however, is not. One can
show that, when going to lower and lower energies, the behaviour of
the system is well described by that of an \emph{effective} system
with $g_{i}\rightarrow g_{i}\left(l\right)$ where the functional
form of the coupling constants is given by the renormalization-group
equations \cite{Gogolin1998}
\begin{align}
\frac{dg_{k}}{dl} & =\left(d-\triangle_{k}\right)-S_{d}\sum_{i\neq j}\left(\triangle_{i}+\triangle_{j}-\triangle_{k}\right)g_{i}g_{j}\label{eq:RG-equations}
\end{align}
where $d$ is the dimensionality and $S_{d}$ is the volume of the
$d$-sphere.

\subsection{Extended RG}



So far, we have neglected the effect of backward (intra-wire) tunnelling processes
in our analysis. Let us therefore add to $H_{\rm 1 P}$ in Eqn.~(\ref{eq:H_single_particle_fermion})
single-particle intra-wire tunnelling terms 
\begin{eqnarray}
\hat{H}_{1P}^{(R)}& = & g_{R}\sum_{j=1,2}\int ds\,\delta\left(s\right)\,\hat{t}_{ j}^{\left(R\right)}(s),
\end{eqnarray}
with 
\begin{eqnarray}
\hat{t}_{ j}^{\left(R\right)}=\hat{\psi}_{R j}^{\dagger}\hat{\psi}_{L j}+{\rm H.c.}
\end{eqnarray}
and to Eqn.~(\ref{eq:H_Interaction_fermionic}) the respective two-particle
intra-wire tunnelling term 
\begin{eqnarray}
\hat{H}_{\rm int}^{\left(R\right)} & = &  G_R\sum_{j=1,2}\,\int ds\,\delta\left(s\right) \hat{V}_{j}^{\left(R\right)}\left(s\right)  
\end{eqnarray}
with
\begin{eqnarray}
\hat{V}_{j}^{\left(R\right)}=\hat{\psi}_{R j}^{\dagger}\hat{\psi}_{L j}\,\hat{\psi}_{R j}^{\dagger}\hat{\psi}_{L j}+ {\rm H.c.} \enspace.
\end{eqnarray}
In bosonized form, these terms read 
\begin{align}
\hat{t}_{j}^{\left(R\right)}\left(s\right) & =\pm_j\frac{\sigma_{z}}{\pi\lambda}\,\sin{\sqrt{4 \pi K}\Phi_{j}\left(s\right)},\\
\hat{V}_{j}^{\left(R\right)}\left(s\right) & =-\frac{1}{2\pi^2\lambda^2}\cos{\sqrt{16\pi K}\,\Phi_{j}\left(s\right)},
\end{align}
and from their autocorrelation function we can determine  their scaling dimension $\triangle_{t_{R}}=K$
and $\triangle_{V_{R}}=4K$ and conformal spin to be zero
- i.e. $\hat{t}_j^{\left(R\right)}$ is \emph{relevant} for $K<1$, \emph{marginal}
for $K=1$ and \emph{irrelevant} for $K>1$, while $\hat{V}_{R}$ is\emph{
relevant }for $K<1/4$, \emph{marginal} for $K=1/4$ and \emph{irrelevant}
for $K>1/4$. 
Performing operator product expansions in standard way  \cite{Gogolin1998}  we arrive at the modified version of our (one-loop) RG-equations
Eq.~(\ref{eq:RG_equations}), with three coupled equations for the
single-particle terms 
\begin{eqnarray}
\frac{dg_{F}}{dl} & = & \left[1-\frac{K+K^{-1}}{2}\right]\,g_{F}-K\,g_{B}g_R\nonumber \\
\frac{dg_{B}}{dl} & = & \left[1-\frac{K+K^{-1}}{2}\right]\,g_{B}-K\,g_{F}g_R\nonumber \\
\frac{dg_R}{dl} & = & \left(1-K\right)\,g_{R}-K^{-1}\,g_{F}g_{B}\label{eq:RG-Equations_Single_particle_High_Symmetry}
\end{eqnarray}
and three coupled equations for the two-particle terms 
\begin{eqnarray}
\frac{dG_F}{dl} & = & \left(1-2K\right)G_F-\left(K^{-1}-K\right)g_{F}^2-g_{R}^2\nonumber \\
\frac{dG_{B}}{dl} & = & \left(1-2K\right)G_{B}-\left(K^{-1}-K\right)g_{B}^2-g_{R}^2\nonumber \\
\frac{dG_{R}}{dl} & = & \left(1-4K\right)G_{R}+Kg_{R}^2\label{eq:RG-Equations_Two_particle_High_Symmetry}
\end{eqnarray}
Note that, as before, the single-particle terms in Eq. (\ref{eq:RG-Equations_Single_particle_High_Symmetry})
contribute to the growth of the two-particle terms Eq. (\ref{eq:RG-Equations_Two_particle_High_Symmetry}),
but not vice versa - when looking for a fixed point in parameter-space,
we must first set the respective r.h.s of Eq. (\ref{eq:RG-Equations_Single_particle_High_Symmetry})
to zero. Integrating Eq. (\ref{eq:RG-Equations_Single_particle_High_Symmetry})
numerically, we find
that that the trivial fixed point $\left(g_{F},g_{B},g_{R}\right)=\left(0,0,0\right)$
is still a fixed point for any $K<1$, but it is no longer a \emph{stable}
fixed point, since $\triangle_{t_{R}}=K$ means that $t_{R}$ is \emph{relevant}
for all $K<1$ - i.e. \emph{only} if the bare value $g_{R}\left(l=0\right)$
is zero can the system flow to the (trivial) fixed point $\left(0,0,0\right)$.
Additionally, due to the one-loop coupling terms (the respective second
terms in the r.h.s of Eqs. (\ref{eq:RG-Equations_Single_particle_High_Symmetry})),
and specifically the term $...-K^{-1}\,g_{F}g_{B}$
in the third equation, the system will also flow \emph{away} from
the trivial fixed point, \emph{even if} $g_{R}\left(l=0\right)=0$
--- unless either $g_{F}\left(l=0\right)=0$ or $g_{B}\left(l=0\right)=0$.

We find that a \emph{finite} fixed point is given for $K=1/2$ and
either $g_{R}\left(l=0\right)=g_{B}\left(l=0\right)=0$, or $g_{R}\left(l=0\right)=g_{F}\left(l=0\right)=0$
 --- where the former case is again the one we have
previously examined. Due to the large momentum-transfer involved in
inter-wire backscattering, it is reasonable to assume that the bare
value of $g_{B}\left(l=0\right)$ will always be smaller than that
of the inter-wire forward-scattering term $g_{F}\left(l=0\right)$,
and hence, $g_{R}\left(l=0\right)=g_{B}\left(l=0\right)=0$ is the
only physically relevant case in which the system flows to a finite
fixed point for repulsive interactions $K=1/2$. In that case, the
system again flows to effective Hamiltonian $H=H_{0}\left[\Phi_{1},\Theta_{1}\right]+H_{0}\left[\Phi_{2},\Theta_{2}\right]+g_{F}\left(l\right)\hat{V}_{F}$ where $H_0$ denotes the Gaussian model.

\setcounter{equation}{0} 
\setcounter{figure}{0}  
\renewcommand{\theequation}{B-\arabic{equation}}  
\renewcommand{\thefigure}{B-\arabic{figure}} 

\section{Charge density waves}\label{AppendixB}
In this Appendix we review the properties of quasi-classical coherent states in quantum mechanics and show how they are related to charge density waves in one-dimensional fermions with linear dispersion relations. We also derive the important Eqn.~(\ref{CDWsol}) in our main text, by deriving an even more general result.

\subsection{The quantum optics of a laser pulse}
Photons in a non-dissipative linear medium are non-interacting bosons, the excitation quanta of the quantized electromagnetic fields. If the fields are decomposed into normal modes, each normal mode is a harmonic oscillator, and a quantum state with $n$ photons in that mode is simply the $n$-th excited state of the quantized oscillator.

If the electromagnetic field is driven by an effectively classical time-dependent charge distribution, then this time-dependent source couples linearly to the field. By spatial Fourier transformation it therefore provides a time-dependent linear drive for every normal mode oscillator. For each normal mode we can write the Hamiltonian
\begin{eqnarray}\label{HHH}
	\hat{H} = \omega\hat{c}^{\dagger}\hat{c} + f(t)\hat{c}^{\dagger} + f^{*}(t)\hat{c}\;,
\end{eqnarray}
where $\omega$ is the normal mode's frequency, $f(t)$ is the spatial Fourier component of the classical source which matches the field's normal mode, and $\hat{c}$ is the canonical lowering operator for the normal mode oscillator, which thus destroys a bosonic photon in this mode of the field.

With $f\to0$ the ground state $|0\rangle$ of $\hat{H}$ is the state annihilated by $\hat{c}$, $\hat{c}|0\rangle =0$, and in general the eigenstates of $\hat{H}$ are $|n\rangle$ such that $\hat{c}^{\dagger}\hat{c}|n\rangle=n |n\rangle$ for any whole number $n$. If this field mode begins at $t=0$ in its ground state, $|\Psi\rangle(0)\rangle=|0\rangle$, but then nonzero $f(t)$ is turned on, the time-dependent Schr\"odinger equation for the evolving quantum state of the mode is solved exactly for any $f(t)$ by
\begin{eqnarray}\label{Glauber0}
	|\Psi(t)\rangle &=& e^{-i\theta(t)}e^{-\frac{1}{2}|\gamma(t)|^{2}}\sum_{n=0}^{\infty}\frac{[\gamma(t)]^{n}}{\sqrt{n!}}|n\rangle\nonumber\\
			\gamma{t} &=& -i\int_{0}^{t}\!dt'\,f(t')e^{-i\omega(t-t')}\nonumber\\
			\theta(t) &=& \frac{1}{2}\int_{0}^{t}\!dt'\,[f(t')\gamma^{*}(t')+f^{*}(t')\gamma(t')]\;.
\end{eqnarray}
The time-dependent drive has a certain time-dependent quantum amplitude to excite any number of photons. The larger $f$ is, and the longer time runs, the larger the amplitudes become to have excited more photons.

The entire class of states of a harmonic oscillator having the form
\begin{eqnarray}\label{Glauber1}
	|\gamma\rangle &\equiv&e^{i\theta}e^{-\frac{1}{2}|\gamma|^{2}}\sum_{n=0}^{\infty}\frac{\gamma^{n}}{\sqrt{n!}}|n\rangle\;,
\end{eqnarray}
for any complex c-number $\gamma$ and real phase $\theta$ are known in quantum optics as \textit{Glauber coherent states} \cite{Glauber}. Our solution (\ref{Glauber0}) to the driven field mode problem means that classically driving an electromagnetic field mode from its ground state produces a particular time-dependent Glauber state. The Glauber coherent states are quasi-classical states; to see this we note that they are eigenstates of the photon destruction operator:
\begin{eqnarray}\label{Glauber2}
	\hat{c}|\gamma\rangle \equiv \gamma |\gamma\rangle\;,
\end{eqnarray}
as is readily seen from the definition (\ref{Glauber1}) and the action $\hat{c}|n\rangle = \sqrt{n}|n-1\rangle$ of the lowering operator. In the particular case (\ref{Glauber0}) where our photon mode has been classically driven with $f(t)$, therefore, we could define a new, time-dependently shifted photon destruction operator,
\begin{eqnarray}\label{}
	\hat{c}' &\equiv& \hat{c}-\gamma(t)\;,
\end{eqnarray}
which would at any time always annihilate the driven quantum state $|\Psi(t)\rangle$. 

Because the shift $\gamma(t)$ is only a c-number, moreover, the commutation relation $[\hat{c}',\hat{c}^{'\dagger}]=1$ of the new destruction operator remains exactly the same as that of the original operator $\hat{c}$. The new operator is therefore every bit as valid as a photon destruction operator as the original one, and by converting our notation to use it, we can say with perfect validity that the driven quantum state remains forever in the ground state, but the destruction operator acquires a time-dependent shift $-\gamma(t)$. We can in fact define an entire new time-dependent basis of $n$-photon states,
\begin{eqnarray}\label{}
	|n\rangle_{\gamma}&\equiv& \frac{[\hat{c}^{\dagger}-\gamma^{*}(t)]^{n}}{\sqrt{n!}}|\gamma(t)\rangle\;,
\end{eqnarray}
effectively redefining photons to be quanta destroyed by $\hat{c}'$ instead of by $\hat{c}$. Using this new, time-dependent basis, it becomes an exactly true statement that the classical source neither creates nor destroys any photons at all, but only shifts the original operators $\hat{c}$ and $\hat{c}^{\dagger}$ by time-dependent c-numbers $\gamma(t)$ and $\gamma^{*}(t)$, respectively:
\begin{eqnarray}\label{}
	\hat{c} = \hat{c}'+\gamma(t)\;,
\end{eqnarray}
where now $\hat{c}'$ is the destroyer of photons.

All of the above is obtained even more straightforwardly if we switch from the Schr\"odinger picture of quantum mechanics, where the quantum states are time-dependent, to the Heisenberg picture of time-independent states and evolving operators. In this representation the Heisenberg equation of motion for $\hat{c}(t)$ under (\ref{HHH}) is
\begin{eqnarray}\label{}
	i\frac{d}{dt}\hat{c} = [\hat{c},\hat{H}] = \omega \hat{c} + f(t)\;.
\end{eqnarray}
If we impose the Heisenberg initial condition at $t=0$, we find exactly $\hat{c}(t)=e^{-i\omega t}\hat{c}(0)+\gamma(t)$. So the Heisenberg initial-time operator $\hat{c}(0)$ is simply our $\hat{c}'$, and the c-number shift $\gamma(t)$ is the exact Heisenberg evolution.

If we repeat the above analysis for every normal mode of the electromagnetic field, we find that Hermitian quantum field components of the form
\begin{eqnarray}\label{}
	\hat{\Phi}(\mathbf{r}) = \sum_{\mathbf{k}}e^{i\mathbf{k}\cdot\mathbf{r}}\hat{c}_{\mathbf{k}}+\mathrm{H.c.}
\end{eqnarray}
are effectively shifted, through the classical sources, by classical fields:
\begin{eqnarray}\label{}
	\hat{\Phi}(\mathbf{r}) &=& \hat{\Phi}'(\mathbf{r}) + \sum_{\mathbf{k}}\left[e^{i\mathbf{k}\cdot\mathbf{r}}\gamma_{\mathbf{k}}(t)+\mathrm{H.c.}\right]\;.
\end{eqnarray}
This is what is meant in quantum optics by saying that a laser pulse---or for that matter a radio broadcast---is a classical electromagnetic field superimposed on the fluctuating quantum vacuum.

\subsection{Charge density waves of one-dimensional chiral fermions}
\subsubsection{Bosonization}
Readers familiar with bosonization for one-dimensional fermions with linear dispersion relations will quickly see how to apply our discussion of photons to the fermionic system. For a single wire the bosonization mapping from the chiral fermionic field $\hat{\psi}_{\alpha}(s)$ to bosonic fields $\hat{\Phi}(s)$ and $\hat{\Theta}(s)$ is
\begin{eqnarray}
\hat{\psi}_{\alpha}(s) & = & \frac{\eta}{\sqrt{2\pi\,\lambda}}e^{i[\pm_{\alpha}\sqrt{\pi K}\,\hat{\Phi}(s)-\sqrt{\pi/K}\,\hat{\Theta}(s)]}\;,\label{Bosonization2}
\end{eqnarray}
where $\alpha=R,L$ denotes right-moving or left-moving fermions, and $\pm_{\alpha}$ is $+$ for right-movers and $-$ for left-movers, respectively. 

Because of the well-known subtlety of normal ordering and the spatial smearing that projects all our fields into the subspace of excitations with the linear dispersion relation, the total charge density at position $s$ is not simply a constant, as one might think naively, but rather proportional to the bosonic field $\hat{\Phi}(s)$:
\begin{eqnarray}\label{}
	\sum_{\alpha=R,L}:\!\hat{\psi}^{\dagger}_{\alpha}(s)\hat{\psi}_{\alpha}(s)\!: = \sqrt{\frac{K}{\pi }}\frac{\partial}{\partial s}\hat{\Phi}(s)\;.
\end{eqnarray}
Since a time-dependent classical voltage applied to the wire couples linearly to the charge density, it will therefore supply linear drives to all the normal modes of the bosonic fields, exactly as in the previous Subsection of this Appendix. 

The result in Heisenberg picture will be to shift the Hermitian bosonic fields by some real classical field, and thereby to multiply the fermionic fields $\hat{\psi}_{\alpha}$ by c-number phases. Again because of the subtlety of spatial smearing and normal ordering, the charge density will still be proportional to the spatial derivative of the bosonic field $\hat{\Theta}(s)$. Since the applied classical voltage has shifted this bosonic field by a certain classical field, the quantum charge density has likewise been shifted by a classical field. The general classical voltage creates classical charge density waves in the system of fermions, and these are represented by c-number phases in the fermionic fields.

\subsubsection{Charge density}
We can confirm that classical phases really do affect the fermionic charge density, even without appealing to bosonization, as long as we remember that the linear dispersion relation which we assume only really applies within a finite range of long wavelengths. We can ignore this, and calculate normally with local quantum field theory, if we in the end project all of our local field operators into the long-wavelength subspace. As long as all the excitations that we actually have are well within this subspace, the precise manner in which we define the projection will not matter. A concrete way of projecting is to smear the spatial arguments of our fields with a narrow Gaussian whose width $\lambda$ defines our short-distance cut-off scale:
\begin{eqnarray}\label{smear}
	``\hat{\psi}(s)" = \frac{1}{\sqrt{2\pi\lambda}}\int\!d\xi\,\hat{\psi}(\xi)e^{-\frac{1}{2\lambda}(\xi-s)^{2}}\;.
\end{eqnarray}
Since we will be considering $\lambda$ to be very short compared to all our excitation wavelengths, for most purposes we will be able to take the limit $\lambda\to0$ and not have to consider the spatial smearing at all. Correctly defining the local charge density, however, is one case in which the smearing does matter.

To see this we can consider a fermionic field operator which has been found, in a local calculation that ignores spatial smearing, to be multiplied by a position-dependent classical phase $\beta(s)$:
\begin{eqnarray}\label{}
	\hat{\psi}_{\beta}(s)=\frac{e^{-i\beta(s)}}{\sqrt{2\pi}}\int\!dk\,\hat{a}_{k}e^{iks}\;.
\end{eqnarray}
Implementing the smearing then means that the charge density, including the contribution from the Fermi sea down $k\sim -1/\lambda$, is
\begin{widetext}
\begin{eqnarray}\label{}
	\hat{\psi}_{\beta}^{\dagger}(s)\hat{\psi}_{\beta}(s) 
&=&\frac{1}{(2\pi)^{2}\lambda^{2}}\int\!dk dk'\,\int\!d\xi d\xi'\,e^{-\frac{1}{2\lambda^{2}}[(\xi-s)^{2}+(\xi'-s)^{2}]}e^{i[\beta(\xi')-\beta(\xi)]}e^{i(k\xi-k'\xi')}\hat{a}^{\dagger}_{k'}\hat{a}_{k}\nonumber\\
&=&\frac{1}{(2\pi)^{2}\lambda^{2}}\int\!dk dk'\,e^{i(k-k')s}\int\!d\xi d\xi'\,e^{-\frac{1}{2\lambda^{2}}(\xi^{2}+\xi^{'2})}e^{i[\beta(s+\xi')-\beta(s+\xi)]}e^{i(k\xi-k'\xi')}\hat{a}^{\dagger}_{k'}\hat{a}_{k}
\;.
\end{eqnarray}

For all $\beta(s)$ that vary slowly compared to the cut-off length $\lambda$, we can expand $\beta(s+\xi)=\beta(s)+\xi\,\beta'(s)+\mathcal{O}(\xi^{2})$ and then perform the $\xi$ and $\xi'$ integrals, obtaining
\begin{eqnarray}\label{consider}
	\hat{\psi}_{\beta}^{\dagger}(s)\hat{\psi}_{\beta}(s) &=& \frac{1}{2\pi}\int\!dk dk'\,e^{i(k-k')s}e^{-\frac{\lambda^{2}}{2}\left([k-\beta'(s)]^{2}+[k'-\beta'(s)]^{2}\right)}\hat{a}^{\dagger}_{k'}\hat{a}_{k}\nonumber\\
&\equiv& \frac{1}{2\pi}\int\!dk dk'\,e^{i(k-k')s}e^{-\frac{\lambda^{2}}{2}\left([k-\beta'(s)]^{2}+[k'-\beta'(s)]^{2}\right)}:\!\hat{a}^{\dagger}_{k'}\hat{a}_{k}\!:  + \frac{1}{2\pi}\int_{-\infty}^{0}\!dk\,e^{-\lambda^{2}[k-\beta'(s)]^{2}}
\end{eqnarray}
when we apply the identity $\hat{a}^{\dagger}_{k'}\hat{a}_{k}\equiv :\!\hat{a}^{\dagger}_{k'}\hat{a}_{k}\!:+\delta(k-k')\theta(-k)$ for the fermionic normal ordering as standardly denoted with $:\dots :$.

For $\beta'(s)$ small compared to $1/\lambda$ we can evaluate the final c-number integral in (\ref{consider}) as
\begin{eqnarray}\label{}
	\frac{1}{2\pi}\int_{-\infty}^{0}\!dk\,e^{-\lambda^{2}[k-\beta'(s)]^{2}} &=&\frac{1}{2\pi}\int_{-\infty}^{-\beta'(s)}\!dk\,e^{-\lambda^{2}k^{2}}\nonumber\\
&\equiv&\frac{1}{2\pi}\int_{-\infty}^{0}\!dk\,e^{-\lambda^{2}k^{2}}-\frac{1}{2\pi}\int_{-\beta'(s)}^{0}\!dk\,e^{-\lambda^{2}k^{2}}\nonumber\\
&=&\frac{1}{\sqrt{2\pi}\lambda} + \beta'(s)\equiv \rho_{0}+\frac{\beta'(s)}{2\pi}
\end{eqnarray}
for $|\beta'(s)|\ll 1/\lambda$. We have recognized $(\sqrt{2\pi}\lambda)^{-1}\equiv\rho_{0}$ as the contribution to charge density of the ground state's filled Fermi sea, when the $k$-space cut-off $1/\lambda$ effectively gives the Fermi sea a finite depth. 

Since we assume that all our excitations will be on wavelengths long compared to $\lambda$, the normally ordered $:\!\hat{a}^{\dagger}_{k'}\hat{a}_{k}\!:$ will simply annihilate all the many-body quantum states that we consider, unless $|k|$ and $|k'|$ are both small compared to $1/\lambda$. We have furthermore assumed that $\beta'(s)$ is small compared to $1/\lambda$. For all the quantum states that we will consider, therefore, we can take the limit $\lambda\to0$ in the term in (\ref{consider}) that is proportional to $:\!\hat{a}^{\dagger}_{k'}\hat{a}_{k}\!:$. We therefore conclude that for the range of system states we consider, the charge density without the Fermi sea contribution is
\begin{eqnarray}\label{CDW1}
	:\!\hat{\psi}_{\beta}^{\dagger}(s)\hat{\psi}_{\beta}(s)\!: & \equiv& \hat{\psi}_{\beta}^{\dagger}(s)\hat{\psi}_{\beta}(s)-\rho_{0}= \frac{1}{2\pi}\int\!dk dk'\,e^{i(k-k')s}:\!\hat{a}^{\dagger}_{k'}\hat{a}_{k}\!:  + \frac{1}{2\pi}\beta'(s)\;.
\end{eqnarray}\end{widetext}
The c-number shift $\beta'(s)/(2\pi)$ in (\ref{CDW1}) shows that the $e^{-i\beta(s)}$ prefactor in $\hat{\psi}$ is really the Heisenberg picture's way of expressing the fact that externally driving the charge density effectively shifts the Fermi level up and down locally. In other words, it excites charge density waves that are truly like surface waves on the Fermi sea.

\subsubsection{Generation of charge density waves}
As we have already indicated in discussing the bosonized representation of charge density waves, the classical charge density waves that are analogous to classical laser fields, and that are represented in Heisenberg picture by time- and space-dependent classical phases multiplying the fermionic field operators, are not only a set of theoretically interesting quantum states. They are also precisely the states which are generated by applying space- and time-dependent classical voltages to a Luttinger liquid. The generation itself is not really relevant to this present paper, since we will simply assume an initial state at $t=0$ in which certain classical charge density waves have already been generated. The fact that external voltages generate precisely this kind of classical charge density waves, however, is what makes this particular kind of initial state experimentally relevant, and not simply a theoretical exercise.

For non-interacting fermions, including the `refermions' of the case $K=1/2$ in this paper, the result that classical driving produces classical phases in the fermionic fields can also be obtained straightforwardly without bosonization, by exploiting the same linear fermionic dispersion relation that makes bosonization work. With an arbitrary time- and space-dependent classical voltage $f(s,t)$, the Hamiltonian for a single wire of free right-moving fermions is
\begin{eqnarray}\label{}
	\hat{H}_{R}&=&\int\!ds\,\left(\hat{\psi}^{\dagger}_{R}(-i\partial_{s})\hat{\psi}_{R} +f(s,t)\hat{\psi}^{\dagger}_{R}\hat{\psi}_{R}\right)\;.
\end{eqnarray}
The Heisenberg equation of motion for the fermionic field operator is then
\begin{eqnarray}\label{Heis2}
	i(\partial_{t}+\partial_{s})\hat{\psi}_{R}(s,t) = f(s,t)\hat{\psi}_{R}(s,t)\;.
\end{eqnarray}
Thanks to the linear dispersion relation, this is a first-order differential equations and thus easy to solve:
\begin{eqnarray}\label{antenna}
	\hat{\psi}_{R}(s,t) &=& e^{-iF_{R}(s,t-t_{I})}\hat{\psi}_{R}(s-t+t_{I},t_{I})\\
	F_{R}(s,t) &=& \frac{1}{2}\int_{s-t}^{s+t}\!du\,f\Big(\frac{u+s-t}{2},\frac{u-s+t}{2}\Big)\nonumber
\end{eqnarray}
relates the time-dependent field to the initial field at $t=t_{I}$, and can be confirmed by straightforward differentiation to solve (\ref{Heis2}). 

We now consider a scenario in which $t_{I}$ is large and negative, and the classical voltage is turned off before $t=0$. If in this scenario we look only at $t>0$, then for all $u>t-t_{I}+s$ the integrand $f\big((u+s-t+t_{I})/2,(u-s+t-t_{I})/2\big)$ has a time argument $(u-s+t-t_{I})/2>t-t_{I}>0$, and therefore vanishes because the classical drive has been turned off long before any positive $t$. For all $t>0$ in this scenario, then, we can replace the upper limit of the $F_{R}$ integral with infinity, and thereby find that $F_{R}(s,t-t_{I}) = \mathcal{A}_{R}(s-t)$ for a certain single-argument function $\mathcal{A}$. 

In a similar way we can conclude for a left-moving fermionic field that once the external driving voltage has been turned off, all the later effects of the driving voltage on $\hat{\psi}_{L}(s,t)$ can be represented by $F_{L}(s,t-t_{I})=\mathcal{A}_{L}(s+t)$ for a certain $\mathcal{A}_{L}$. 

\subsection{Charge density waves in the $K=1/2$ Luttinger intersection}

We are now ready to determine the combined effects in our two-wire $K=1/2$ Luttinger system of both driving to generate incoming charge density waves, and propagation through the intersection, by adding driving voltages $f_{1,2}(s,t)$ on each wire to our refermionized Hamiltonian (\ref{eq:Refermionized_Hamiltonian}):\begin{widetext}
\begin{eqnarray}\label{drive}
	\hat{H}\to\hat{H}_{f}& = &  \sum_{\pm}\int ds\,\left\{ :\hat{\Psi}_{R\pm}^{\dagger}\left(-i\,\partial_{s}\right)\hat{\Psi}_{R\pm}:+:\hat{\Psi}_{L\pm}^{\dagger}\left(+i\,\partial_{s}\right)\hat{\Psi}_{L\pm}:\right\}  -2i\,V\int\!ds\,\delta(s)\,\left\{ \,\hat{\Psi}_{R+}^{\dagger}\hat{\Psi}_{L+}\left(s\right)-\hat{\Psi}_{L+}^{\dagger}\hat{\Psi}_{R+}\left(s\right)\right\}\nonumber\\
&& \qquad + \int\!ds\,\left(f_{1}(s,t)\hat{\rho}_{1}(s)+f_{2}(s,t)\hat{\rho}_{2}(s)\right)\;.
\end{eqnarray}
\subsubsection{`Minus' fields}
We begin by looking at the simpler part of our system, namely the $\hat{\Psi}_{\alpha-}$ fields, which are not affected by the intersection. Inserting (\ref{density0}) into (\ref{drive}) reveals that for the $\hat{\Psi}_{\alpha-}$ fields the Heisenberg equations of motion under $H_{f}$ are
\begin{eqnarray}\label{HEOMdriven-}
	i\left(\frac{\partial}{\partial t}\pm_{\alpha}\frac{\partial}{\partial s}\right)\hat{\Psi}_{\alpha-}(s,t)&=&\frac{1}{2}[f_{1}(s,t)-f_{2}(s,t)]\hat{\Psi}_{\alpha-}(s,t)\;.
\end{eqnarray}

The solutions to (\ref{HEOMdriven-}) with the initial condition $\hat{\Psi}_{\alpha-}(s,t_{I})=\hat{\Psi}_{\alpha-}(s)$ can easily be confirmed by straightforward differentiation to be
\begin{eqnarray}\label{PsiMinusSol}
		\hat{\Psi}_{\alpha-}(s,t)&=&\frac{e^{-iF_{\alpha-}(s,t-t_{I})}}{\sqrt{2\pi}}\int\!dk\,e^{ik(\pm_{\alpha}s-t+t_{I})}\hat{a}_{\alpha k-}\nonumber\\
		F_{R\pm}(s,t)&=& \frac{1}{2}\int_{s-t}^{s+t}\!du\,[f_{1}\Big(\frac{u+s-t}{2},\frac{u-s+t}{2}\Big)\pm f_{2}\Big(\frac{u+s-t}{2},\frac{u-s+t}{2}\Big)]\nonumber\\
		F_{L\pm}(s,t)&=&\frac{1}{2}\int_{s-t}^{s+t}\!du\,[f_{1}\Big(\frac{u+s+t}{2},\frac{s+t-u}{2}\Big)\pm f_{2}\Big(\frac{u+s+t}{2},\frac{s+t-u}{2}\Big)]\;,
\end{eqnarray}
where the $\hat{a}_{\alpha k-}$ operators are the ones from (\ref{expand}) that diagonalize $H$ without the $f_{j}$ driving. The $F_{\alpha+}$ fields will appear in the more complicated solutions for $\hat{\Psi}_{\alpha+}(s,t)$ that we will find below.

Just as we saw in the previous Subsection, if the driving $f_{j}$ turn off before $t=0$, then after $t=0$ we can write $F_{R\pm}(s,t-t_{I})=\mathcal{A}_{R\pm}(s-t)$ and $F_{L\pm}(s,t-t_{I})=\mathcal{A}_{L\pm}(s+t)$. We can also freely absorb the $s$- and $t$-independent phases $e^{ikt_{I}}$ into the $\hat{a}_{\alpha k-}$ operators, since this phase multiplication alters neither the anti-commutation relations of the $\hat{a}_{\alpha k-}$ operators nor their diagonalization of $\hat{H}$. We therefore obtain from (\ref{PsiMinusSol}) the solutions for $\hat{\Psi}_{\alpha-}(s,t)$ that are given in the first half of (\ref{CDWsol}) of our main text.

\subsubsection{`Plus' fields}
We now turn to the more complicated problem of the $\hat{\Psi}_{\alpha+}(s,t)$. Their Heisenberg equation of motion with driving included reads
\begin{eqnarray}\label{HEOMdriven+}
	i\left(\frac{\partial}{\partial t}+\frac{\partial}{\partial s}\right)\hat{\Psi}_{R+}(s,t)&=&-2iV\delta(s)\hat{\Psi}_{L+}(0,t)+\frac{1}{2}[f_{1}(s,t)+f_{2}(s,t)]\hat{\Psi}_{R+}(s,t)\nonumber\\
	i\left(\frac{\partial}{\partial t}-\frac{\partial}{\partial s}\right)\hat{\Psi}_{L+}(s,t)&=&+2iV\delta(s)\hat{\Psi}_{R+}(0,t)+\frac{1}{2}[f_{1}(s,t)+f_{2}(s,t)]\hat{\Psi}_{L+}(s,t)\;.
\end{eqnarray}
These coupled equations are less trivial than (\ref{HEOMdriven-}), but they are still a linear system of first-order differential equations that can be solved exactly. If we already in advance absorb the time- and space-independent phases $e^{ikt_{I}}$ into the $\hat{a}_{\alpha k+}$ operators as we did above with their $\hat{a}_{\alpha k-}$ analogs, the solutions which apart from the absorbed $e^{ikt_{I}}$ factors satisfy the initial condition $\hat{\Psi}_{\alpha+}(s,t_{I})=\hat{\Psi}_{\alpha+}(s)$ are
\begin{eqnarray}\label{RTsol}
\hat{\Psi}_{R+}(s,t)&=&\frac{e^{-iF_{R+}(s,t-t_{I})}}{\sqrt{2\pi}}\int\!dk\,e^{ik(s-t)}\left([\theta(-s)+\mathcal{T}\theta(s)]\hat{a}_{Rk+}
				- \mathcal{R}\theta(s) e^{i\Delta F(t-t_{I}-s)}\hat{a}_{Lk+}\right)\\ 
\hat{\Psi}_{L+}(s,t)&=&\frac{e^{-iF_{L+}(s,t-t_{I})}}{\sqrt{2\pi}}\int\!dk\,e^{-ik(s+t)}\left([\theta(s)+\mathcal{T}\theta(-s)]\hat{a}_{Lk+}
				+ \mathcal{R}\theta(-s) e^{-i\Delta F(t-t_{I}+s)}\hat{a}_{Rk+}\right)\nonumber\\
	\Delta F(t) &\equiv& [F_{R+}(0,t)-F_{L+}(0,t)]\theta(t)\;.
\end{eqnarray}
Despite the apparent complexity of these solutions it is straightforward to confirm them by differentiating. \end{widetext}

If we now consider as before that the driving voltages which began at the large negative $t_{I}$ are turned off before $t=0$, for $t>0$ we can again write $F_{\alpha+}(s,t)=\mathcal{A}_{\alpha}(s\mp_{\alpha} t)$. The new phase $\Delta F$ must also be considered, however. It contains $F_{\alpha+}$ functions with time arguments $t-t_{I}\mp_{\alpha}s$, and for large enough $|s|>t-t_{I}$ these time arguments may not be greater than zero. If $t-t_{I}\mp_{\alpha}s < 0$, then in fact the step function in the definition of $\Delta F$ will make $\Delta F$ vanish, instead being equal to any non-vanishing $\mathcal{A}_{\alpha+}$ function. 

There is no need for us to consider any experimental measurements at positions $|s|>t-t_{I}$, however. For all positive $t$ such positions are very far from the intersection (recall that $t_{I}$ is large and negative). They are so far from the origin, in fact, that there has not been time since $t_{I}$ for any signal to have reached them from the intersection, let alone to have reached them after passing through the intersection from some more distant starting point. Measurements at this remote locations will therefore show no effects at all from the intersection, and we can ignore them.

Once we restrict our attention to $|s|<t-t_{I}$ as well as $t>0$, we see that $\Delta F(t-t_{I}\mp s) = \mathcal{A}_{R+}(-t \pm s )-\mathcal{A}_{L+}(t\mp s)$. Inserting this simplification into (\ref{RTsol}) above completes the derivation of the all-important Eqn.~(\ref{CDWsol}) of our main text.

\end{document}